\documentclass[12pt]{article}

\usepackage{amsfonts}
\usepackage{bbm}
\usepackage{cite}
\usepackage{epsf}

\newcommand{\bmat}{\left(\begin{array}}
\newcommand{\emat}{\end{array}\right)}

\def\gtrsim{\mathrel{\raise.3ex\hbox{$>$\kern-.75em\lower1ex\hbox{$\sim$}}}}

\def\p{\partial}
\def\a{\alpha}

\def\b{\beta}
\def\g{\gamma}

\def\th{\theta}
\def\om{\omega}
\def\Om{\Omega}

\def\-{\hphantom{-}}
\def\ov{\overline}
\def\s2{\frac{1}{\sqrt2}}

\def\oh{\frac{1}{2}}
\def\beq{\begin{equation}}
\def\eeq{\end{equation}}
\def\beqa{\begin{eqnarray}}
\def\eeqa{\end{eqnarray}}
\def\D{{\rm D}}

\def\im{{\rm Im \,}}
\def\re{{\rm Re \,}}
\def\tr{{\rm tr \,}}
\def\Tr{{\rm Tr \,}}
\def\dim{{\rm dim \,}}

\def\T{{\bf T}}
\def\TT{{\bf T}}

\def\eps{\epsilon}

\def\CD {{\cal D}}

\def\CM {{\cal M}}
\def\CQ {{\cal Q}}
\def\CR {{\cal R}}
\def\CN {{\cal N}}
\def\CF {{\cal F}}

\def\CS {{\cal S}}
\def\CV {{\cal V}}
\def\cc{{\cal C}}

\def\eps{{\epsilon }}

\def\cn{{\mathcal N}}

\def\IN{\mathbb{N}}
\def\IZ{\mathbb{Z}}
\def\IR{\mathbb{R}}
\def\IP{\mathbb{P}}

\def\mg{m_{3/2}}
\def\mg2{m^2_{3/2}}

\def\deq#1{\mbox{$D$=#1}}
\def\neq#1{\mbox{$\cn$=#1}}
\def\Dsl{\,\raise.15ex\hbox{/}\mkern-13.5mu D} 

\newcommand{\drawsquare}[2]{\hbox{%
\rule{#2pt}{#1pt}\hskip-#2pt
\rule{#1pt}{#2pt}\hskip-#1pt
\rule[#1pt]{#1pt}{#2pt}}\rule[#1pt]{#2pt}{#2pt}\hskip-#2pt
\rule{#2pt}{#1pt}}

\newcommand{\fund}{\raisebox{-.5pt}{\drawsquare{6.5}{0.4}}}
\newcommand{\Ysymm}{\raisebox{-.5pt}{\drawsquare{6.5}{0.4}}\hskip-0.4pt%
        \raisebox{-.5pt}{\drawsquare{6.5}{0.4}}}
\newcommand{\Yasymm}{\raisebox{-3.5pt}{\drawsquare{6.5}{0.4}}\hskip-6.9pt%
        \raisebox{3pt}{\drawsquare{6.5}{0.4}}}
\newcommand{\antifund}{\overline{\fund}}

\begin{document}
\pagestyle{plain}

\makeatletter
\@addtoreset{equation}{section}
\makeatother
\renewcommand{\theequation}{\thesection.\arabic{equation}}
\pagestyle{empty}
\rightline{IFT-UAM/CSIC-06-34}
\rightline{LMU-ASC 51/06}
\begin{center}
\LARGE{Coisotropic D8-branes and Model-building 
\\[10mm]}
\large{ A. Font$^*$\footnote{On leave from Departamento de F\'{\i}sica, Facultad de Ciencias,
Universidad Central de Venezuela, A.P. 20513, Caracas 1020-A, Venezuela.},
 L.E. Ib\'a\~nez$^*$ and F. Marchesano $^{**}$ \\[6mm]}
\small{
* Departamento de F\'{\i}sica Te\'orica C-XI
and Instituto de F\'{\i}sica Te\'orica  C-XVI,\\[-0.3em]
Universidad Aut\'onoma de Madrid,
Cantoblanco, 28049 Madrid, Spain 
\\[1mm]} 
\small{
** Arnold-Sommerfeld-Center for Theoretical Physics,\\
Department f\"ur Physik, Ludwig-Maximilians-Universit\"at M\"unchen,\\
Theresienstra\ss e 37, D-80333 M\"unchen, Germany\\[-0.3em]
}

\vspace*{1cm}

\small{\bf Abstract} \\[1mm]
\end{center}
{\small  
Up to now chiral type IIA vacua have been mostly based on intersecting D6-branes wrapping special Lagrangian 3-cycles on a ${\bf CY}_3$ manifold. We argue that there are additional BPS D-branes which have so far been neglected, and which seem to have interesting model-building features. They are coisotropic D8-branes, in the sense of Kapustin and Orlov. The D8-branes wrap 5-dimensional submanifolds of the ${\bf CY}_3$ which are trivial in homology, but contain a worldvolume flux that induces D6-brane charge on them. This induced D6-brane charge not only renders the D8-brane BPS, but also creates $D=4$ chirality when two D8-branes intersect. 
We discuss in detail the case of a type IIA $\TT^6/(\IZ_2 \times \IZ_2)$ orientifold, where we provide explicit examples of coisotropic D8-branes. We study the chiral spectrum, SUSY conditions, and effective field theory of different systems of D8-branes in this orientifold, and show how the magnetic fluxes generate a superpotential for untwisted K\" ahler moduli. Finally, using both D6-branes and coisotropic D8-branes we construct new examples of MSSM-like type IIA vacua.}

\newpage
\setcounter{page}{1}
\pagestyle{plain}
\renewcommand{\thefootnote}{\arabic{footnote}}
\setcounter{footnote}{0}

\section{Introduction}
\label{sec:intro}

One of the most popular techniques to obtain $D=4$ chiral string compactifications is the construction of type IIA orientifolds with D6-branes intersecting at angles \cite{bgkl00,afiru00} (see \cite{reviews} for reviews on this subject). The building blocks in these constructions are BPS D6-branes, wrapping 3-cycles corresponding to special Lagrangian submanifolds in a ${\bf CY}_3$. In this way one can obtain semirealistic three generation models with a low-energy spectrum quite close to that of the MSSM. Remarkably simple and successful are the intersecting brane models based on the $\T^6/\IZ_2\times \IZ_2$ orientifold \cite{csu01}. In particular, it was shown in \cite{ms04} that a simple local set of D6-branes leading to a MSSM-like spectrum \cite{cimDESY,cimyuk} can be simply embedded into this orientifold background.

Here we would like to point out that, while the above type IIA picture is rather compelling, it 
is in fact far from complete. In particular, we will show that there are other BPS D-branes in these backgrounds, 
and whose model building applications have been ignored up to now. These are nothing but BPS D8-branes, wrapping coisotropic 5-cycles in the ${\bf CY}_3$ and with a non-trivial magnetic flux in their internal worldvolume. 

Coisotropic A-branes were first introduced by Kapustin and Orlov \cite{ko01} in the context of topological string theory, and seem to be required for a complete formulation of the Homological Mirror Symmetry conjecture. Here we will be interested in the particular case of D8's wrapping coisotropic 5-cycles in a {\bf CY}$_3$. That such D$p$-brane exists may look surprising at first sight, since all 5-cycles in a generic {\bf CY}$_3$ are homologically trivial and one would not expect a stable object constructed out of them. However, as we will discuss in detail, coisotropic D8-branes are not only stable but also BPS, and this is because the magnetic flux on the D8-brane induces a non-trivial D6-brane charge on its worldvolume. 

Coisotropic D8-branes have several interesting properties. Chirality arises in D8-D8 or D8-D6 systems due to a mixture of intersecting/magnetization chirality mechanisms. Because of this, Yukawa couplings among chiral fields are generated by a combination of wavefunction overlapping and open string world-sheet instantons. In addition, the BPS conditions on the coisotropic branes give rise to constraints which in the effective field theory may be interpreted as F and D-term cancellation. We will explicitly compute such F and D-terms in the particular case of the $\T^6/\IZ_2 \times \IZ_2$ orientifold, where most of our discussion will be based. Unlike in the case of D6-branes at angles we will see that the F-term generated for D8-branes is non-trivial, and that it involves the untwisted K\"ahler moduli of the compactification. 

From the model-building point of view these D8-branes also present a number of advantages over analogous models with only D6-branes at angles. In particular, the D6-charge carried by coisotropic D8-branes is not of the form {\small (3-cycle) $=$ (1-cycle) $\times$ (1-cycle) $\times$ (1-cycle)}, but rather a sum of two of these. This adds flexibility to the model-building and allows for possibilities not available with standard D6-branes, as we will show by means of explicit examples. Finally, the presence of D8-brane generates a superpotential for the K\"ahler moduli and some open string moduli of our compactification, without the need of closed string NSNS or RR fluxes. Notice that the constructions here discussed are CFT's, unlike the compactifications in which closed string fluxes are added. 

The structure of the paper is as follows. In the next section the notion of coisotropic D$p$-branes is reviewed 
and we argue why there should exist coisotropic D8-branes in generic {\bf CY}$_3$'s. In Section \ref{coisorientifolds} 
we discuss in detail the cases where our compact manifold is $\T^6$ and its $\IZ_2 \times \IZ_2$ orbifold, 
constructing explicit examples of BPS D8-branes on both backgrounds. We analyze the constraints coming from 
the BPS conditions of such D-branes and derive the RR tadpole cancellation conditions. In Section \ref{chirality} 
we discuss how chiral fermions appear at the overlap of D8-branes with some other D8 or D6-branes, and show 
that their multiplicity can be computed as the intersection number of two 3-cycles. Some aspects of the 
$\CN=1$ effective field theory of coisotropic D8-branes are examined in Section \ref{effective}, including a 
discussion on the gauge kinetic function, F- and D-terms, massive $U(1)$'s and Yukawa couplings. 
Section \ref{modelitos} illustrates possible model-building applications of coisotropic D8-branes. 
In particular we present two 3-generation MSSM-like models based on  both coisotropic D8-branes and intersecting D6-branes. Finally, in section \ref{mirror} we analyze the T-duals of the present constructions, and in particular the mirror type I vacua where coisotropic D8-branes become tilted D5-branes or D9-branes with off-diagonal magnetic fluxes. Some final comments are left for Section \ref{tachan}, whereas in Appendix \ref{app} we briefly review the formal definition of coisotropic branes.

\section{Coisotropic D8-branes}

Let us consider type IIA string theory compactified on a Calabi-Yau three-fold $\CM_6$. If we are interested in obtaining a $D=4$ semi-realistic effective theory, we need to embed the Standard Model gauge group and matter content in our construction. This cannot be achieved by just considering closed strings \cite{dkv87}, but we need to include space-filling D-branes in this setup. If, in addition, we aim to construct an $\CN=1$ effective field theory, we need to consider BPS, space-filling D-branes.

In principle, type IIA theory contains three different kinds of D-branes which may be both space-filling and BPS: D4, D6 and D8-branes. However, if our compactification manifold $\CM_6$ has proper $SU(3)$ holonomy, one would never think of obtaining a BPS D-brane out of a space-filling D4-brane.  The reason is that for such manifolds $b_1 = 0$, and then our D4-brane (which is wrapping a 1-cycle of $\CM_6$) would couple to a RR-potential $C_5$ without zero modes. This implies that the Chern-Simons action for a space-filling D4-brane vanishes, and so D4-branes wrapped on 1-cycles cannot carry any central charge. In practice, this is summarized by stating that D4-branes wrap 1-cycles which are `homologically trivial'.\footnote{Such terminology is somewhat misleading, because a D4-brane could still be wrapping a torsional 1-cycle of $\CM_6$.}

The situation is quite different for D6-branes wrapping 3-cycles of $\CM_6$. Here, because $b_3(\CM_6) \not= 0$, the Calabi-Yau background gives rise to a plethora of central charges, which are classified by the elements of $H^3(\CM_6, \IR)$. The BPS conditions for such D6-brane can be expressed in terms of the holomorphic and K\"ahler forms $\Om$ and $J$ \cite{mmms99}, and read
\beq
\begin{array}{rcl}
\Om & = & {\sqrt{|G + \CF|} \over \sqrt{|G|}}\, d{\rm vol}_{3}\\
\CF + i J & = & 0
\end{array}
\label{SUSYD6}
\eeq
where the background forms $\Om$, $J$ and the metric $G$ are pull-backed to the 3-cycle $\Pi_3$ wrapped by the D6-brane. In the language of special holonomy manifolds, these two conditions mean that $\Pi_3$ is a special Lagrangian submanifold of $\CM_6$, with a flat bundle $\CF = 2\pi \a' F + B$. Finally, from the point of view of the $\deq 4$ gauge theory arising from the D6-brane, these BPS conditions can be rephrased as D-flatness and F-flatness equations. More precisely
\beq
\begin{array}{ccrcl}
{\rm D-flatness} & \quad \quad & \im \Om & = & 0\\
{\rm F-flatness} & \quad \quad & \CF + i J & = & 0
\end{array}
\label{SUSYD6b}
\eeq

One would naively say that D6-branes wrapping special Lagrangian 3-cycles exhaust all the possibilities of space-filling BPS D-branes. Indeed, we have already seen that we cannot construct them from D4-branes. A similar argument seems to hold for D8-branes, which also wrap 5-cycles $\Pi_5 \subset \CM_6$ trivial in homology. However, this is not quite true because, unlike the D4-brane, a D8-brane is allowed to carry a non-trivial gauge bundle $F$ without breaking Poincar\'e invariance. This gauge bundle modifies the Chern-Simons action of the D8-brane and, in particular, induces a D6-brane charge on its worldvolume \cite{douglas95} (see also \cite{aads00}). If such D6-brane charge corresponds to a non-trivial element of $H^3(\CM_6, \IR)$, then we can have a non-trivial Chern-Simons action via the coupling
\beq
\int_{M_4 \times \Pi_5} \CF \wedge C_7
\label{CSD8}
\eeq
and thus our D8-brane will have a non-trivial central charge, again related to $H^3(\CM_6, \IR)$. Hence, we should be able to find a BPS stable object.

Given our past experience with type IIA orientifold vacua it may seem quite striking that, besides D6-branes wrapping special Lagrangians, there could also exist stable D8-branes which are mutually BPS with the former. However, this possibility has already arisen in the quite different context of topological string theory. Indeed, as described in \cite{witten92} a D-brane wrapping a special Lagrangian (or rather just Lagrangian) $n$-cycle of a Calabi-Yau $n$-fold is the prototypical example of D-brane in the topological A-model, usually dubbed as A-brane. Naively, these are the only boundary conditions which are allowed for open strings in the A-model but, as shown by Kapustin and Orlov in \cite{ko01}, this is in fact not true. An A-brane does not necessarily wrap a Lagrangian $n$-cycle, but can also wrap a coisotropic $(n+2k)$-cycle, $k = 1, \dots, [n/2]$, if the appropriate worldvolume bundle $F$ is introduced.\footnote{See the Appendix A for a mathematical definition of coisotropic submanifold.} As emphasized by the authors, this fact proves essential in order to understand the full spectrum of D-branes in our theory and, in more theoretical grounds, to correctly formulate the Homological Mirror Symmetry conjecture.\footnote{See \cite{gm04} for a recent perspective of this problem in terms of pure spinors.}

All this suggests that our previous BPS D8-brane should in fact correspond to a coisotropic D8-brane, in the sense of Kapustin and Orlov. Let us further motivate this by analyzing the supersymmetry conditions for coisotropic D-branes, which can be obtained either by a worldsheet \cite{ko01,kl03} or a worldvolume approach \cite{ms05}. In the case of a D8-brane on a Calabi-Yau three-fold they can be written as 
\beq
\begin{array}{rcl}
\CF \wedge\Om  & = & {\sqrt{|G + \CF|} \over \sqrt{|G|}}\, d{\rm vol}_{5}\\
(\CF + i J)^2 & = & 0
\end{array}
\label{SUSYD8}
\eeq
where $\Om$, $J$ and $G$ are now pull-backed over the 5-cycle $\Pi_5$. Notice that, when written in this way, the BPS conditions for D6 and D8-branes look extremely similar.

The D8-brane BPS equations take a particularly suggestive form when the pull-back of $B$ vanishes identically, and then $\CF = 2\pi \a' F$ defines a homology class of 3-cycles $[\Pi_3^F]$ via Poincar\'e duality in $\Pi_5$. This homology class is nothing but the D6-brane charge induced by the Chern-Simons coupling (\ref{CSD8}), which can now be written as
\beq
\mu_8 \int_{M_4 \times \Pi_5} 2\pi \a' F \wedge C_7\, =\, \mu_6 \int_{M_4 \times \Pi_3^F} C_7
\label{CSD8b}
\eeq
where $\mu_p = (2\pi)^{-p} \a'^{-\frac{p+1}{2}}$, and
$\Pi_3^F$ is any representative of $[\Pi_3^F] $ (the Poincar\'e dual of  $[F/2\pi]$). 
Because $[\Pi_3^F]$ is the central charge carried by our D8-brane, it should match the D8-brane tension 
in the BPS limit. This is indeed the case, since by integrating the first equation in (\ref{SUSYD8}) we obtain
\beq
T_{D8} \, =\,  \mu_8 \int_{M_4 \times \Pi_5} 2\pi \a' F \wedge \Om \, =\, \mu_6 \int_{M_4 \times \Pi_3^F} \Om
\label{BPSD8}
\eeq
and so the whole tension of the D8-brane equals that of a D6-brane wrapping a special Lagrangian in $[\Pi_3^F]$, matching the r.h.s. of eq.(\ref{CSD8b}). 

Regarding the second supersymmetry condition, it implies that $F \wedge J = 0$, and hence that $\Pi_3^F$ cannot contain any holomorphic non-vanishing 2-cycle. This is automatic for Lagrangian 3-cycles, which supports the idea that $\Pi_3^F$ could be seen as a special Lagrangian 3-cycle in $\Pi_5$. In addition, (\ref{SUSYD8}) implies that $\CF^2 = J^2$ on $\Pi_5$, and this suggests that the D8-brane charge and the induced D4-brane charge of a coisotropic D8-brane, measured as
\beq
\mu_8 \int_{M_4 \times \Pi_5} C_9 \quad \quad {\rm and} \quad \quad \mu_8 \int_{M_4 \times \Pi_5} \CF^2 \wedge C_5
\label{CSD8l}
\eeq
need to be equal in magnitude.

Finally, we can again rewrite our BPS conditions in terms of $D=4$ D-flatness and F-flatness conditions, namely \cite{kl03,martucci06}
\beq
\begin{array}{ccrcl}
{\rm D-flatness} & \quad \quad & \im (\CF \wedge \Om) & = & 0\\
{\rm F-flatness} & \quad \quad & (\CF + i J)^2 & = & 0
\end{array}
\label{SUSYD8b}
\eeq
While the D-flatness condition of an A-brane is exact at all orders in $\a'$, in general we would expect to receive $\a'$ corrections to the F-flatness condition. This is indeed the case for special Lagrangian D6-branes, where the corrections arise from open string world-sheet instantons \cite{openws}. Although these corrections are less understood in the case of coisotropic A-branes, a proposal in terms of generalized Floer homology has been given in \cite{ko03}.

To summarize, we have argued that D6-branes are not the only space-filling BPS objects in type IIA Calabi-Yau compactifications, and that BPS D8-branes can also exist when endowed with the appropriate worldvolume flux $F$. We have identified such D8-branes as coisotropic A-branes, and have used their supersymmetry conditions to further motivate our initial intuition. Even in the coisotropic D-brane literature, the fact that  $b_5 = 0$ in most Calabi-Yau manifolds has led to believe that stable A-branes of this kind do not exist. However, as our previous discussion shows, the fact that $\Pi_5$ is homologically trivial should not be an obstruction to construct a BPS object. This can still be achieved if  the D-brane charge induced by $F$ is non-trivial in the homology of $\CM_6$. In the next section, we will give further evidence of this general picture, by explicitly constructing BPS D8-branes in Calabi-Yau manifolds with proper $SU(3)$ holonomy.

\section{Coisotropic D8-branes in type IIA orientifolds \label{coisorientifolds}}

In this section we provide explicit examples of coisotropic D8-branes. This will not only illustrate the discussion above, but also prepare the ground to build $\CN=1$ vacua based on D8-branes. We will consider two different backgrounds, which are $\T^6$ and $\T^6/\IZ_2 \times \IZ_2$. Eventually, we will also need to cancel the D8-brane charge and tension without breaking supersymmetry. This can be achieved by introducing O6-planes in our compactification, that is by applying the usual type IIA orientifold projection on the above backgrounds.

\subsection{D8-branes on $\T^6$}

There are not many examples of coisotropic D-branes in the literature. Most of the constructions are based on compactifications on $\T^4$ \cite{ko01,az05} and {\bf K3} \cite{kw06}. Because these are 4-dimensional manifolds, the coisotropic D-brane wraps the whole manifold and hence a non-trivial 4-cycle. The situation is more involved for a D8-brane on a Calabi-Yau three-fold, because in general all the 5-cycles on $\CM_6$ will be trivial. Two exceptions are $\T^6$ and $\T^2 \times {\bf K3}$, where coisotropic D8-branes are known to exist. Again, the known examples are very few, so one of our goals will be to classify the set of coisotropic D8-branes in these backgrounds. We will now proceed to analyze the case of $\T^6$, while $\T^2 \times {\bf K3}$ will arise later on, when considering D8-branes on $\T^6/\IZ_2 \times \IZ_2$. 

Let us then consider type IIA theory compactified on a toroidal background of the form $(\T^2)_1 \times (\T^2)_2 \times (\T^2)_3$, where each two-torus has a rectangular shape. This family of manifolds is parameterized by six compactification radii, which can be arranged as three K\"ahler and three complex structure moduli. The complex structure moduli $i\tau_j$ are pure imaginary quantities, and they enter into the holomorphic 3-form as
\beqa
\Om & = & (4\pi^2\a')^{3/2}\,  \prod_j \left(\frac{\re T_j}{\tau_j}\right)^{1/2} \, dz^1 \wedge dz^2 \wedge dz^3
\label{holo1} \\
dz^j & = & dx^j + i \tau_j dy^j
\label{holo2}
\eeqa
where $x^j \sim x^j +1$ and $y^j \sim y^j +1$ are periodic coordinates, and $\re T_j$ is the area of the $j^{th}$ $\T^2$, in units of $4\pi^2\a'$. This K\"ahler modulus is complexified by including the background B-field on $(\T^2)_j$, so that the combination $4\pi^2 \a'\ T_j = A_j + i B_j$ enters the complexified K\"ahler form $J_c$ as\footnote{In our conventions $\int_{\T^6}  dx^1 \wedge dx^2 \wedge dx^3 \wedge dy^1 \wedge dy^2 \wedge dy^3 = 1$, so that $J^3$ measures the volume of $\T^6$. \label{conv}}
\beq
J_c\,  =\,  B + iJ\, =\, 
4\pi^2\a' \sum_{j=1}^3 {T_j \over 2 \tau^j}\, dz^j \wedge d\bar{z}^j 
\, =\, i\, 4\pi^2 \a' \sum_{j=1}^3 T_j\, dy^j \wedge dx^j.
\label{kahler}
\eeq
Of course, this is just a subspace of the whole $\T^6$ moduli space but, once the $\IZ_2 \times \IZ_2$ and orientifold projection are implemented, any other $\T^6$ modulus will be projected out. We will hence restrict ourselves to the above setup.

\begin{figure}
\epsfysize=4cm
\begin{center}
\leavevmode 
\epsffile{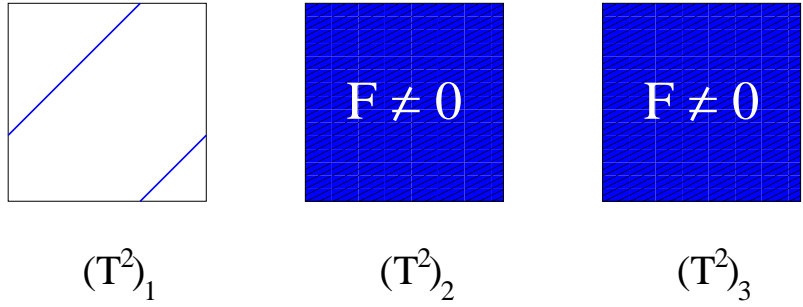}
\end{center}
\caption{Coisotropic D8-brane on $\T^6$.}
\label{coiso1}
\end{figure}

Any 5-cycle of our compactification is of the form $\Pi_5 = \T^5 = {\bf S}^1 \times \T^4$, so our D8-brane will be wrapping two complex dimensions and a 1-cycle in the third complex dimension. We will denote such 5-cycle as
\beq
\Pi_5\, =\, (n^i,m^i)_i \times (\T^2)_j \times (\T^2)_k, \quad \quad i\not= j\not= k\not= i
\label{5cycle} 
\eeq
where $n^i, m^i \in \IZ$ are the D8-brane wrapping numbers along the directions $x^i$ and $y^i$, respectively, and $\{ijk\}$ are in cyclic permutation of $\{123\}$. We now construct a coisotropic D8-brane by introducing a non-trivial gauge bundle $F =dA$ on the four-torus $(\T^2)_j \times (\T^2)_k$, such that the supersymmetry conditions (\ref{SUSYD8}) are satisfied (see figure \ref{coiso1}). For instance, let us a consider a single D8-brane such that
\beq
\begin{array}{rcl}
\Pi_5 & = & (1,0)_1 \times (\T^2)_2 \times (\T^2)_3\\
F/2\pi & = & dx^2 \wedge dx^3 - dy^2 \wedge dy^3
\end{array}
\label{cc}
\eeq
A simple computation gives
\beq
\begin{array}{rcl}
F^2 & = & 4\pi^2\, dx^2 \wedge dy^2 \wedge dx^3 \wedge dy^3 \\
J_c \wedge F & = &  0\\
J_c^2 & = & - (4\pi^2\a')^2\, T_2 T_3\, dx^2 \wedge dy^2 \wedge dx^3 \wedge dy^3
\end{array}
\label{Ftermcc}
\eeq
where all powers of $J_c$ have been pull-backed to $\Pi_5$. The second supersymmetry condition in (\ref{SUSYD8}) then reads
\beq
(\CF + i J)^2\, =\, (2\pi \a' F + J_c)^2 \, =\, 0 \quad \iff \quad T_2T_3 \, =\, 1
\label{Ftermcc2}
\eeq

On the other hand we have that
\beq
\CF \wedge \Om = (4\pi^2\a')^{5/2}\, \prod_j \left(\frac{\re T_j}{\tau_j}\right)^{1/2} (1 + \tau_2 \tau_3) \, {d{\rm vol}_{\Pi_5} \over \sqrt{G}}
\label{Dtermcc}
\eeq
where again $\Om$ and the metric $G$ are pull-backed to $\Pi_5$. This quantity needs to be proportional to the D8-brane DBI action, whose square is given by
\beqa\nonumber
\det \left( G + \CF \right) & = & (4\pi^2\a')^5\, \frac{\re T_1}{\tau_1}\, 
\left[ |T_2 T_3 -1|^2 + {\re T_2 \re T_3 \over \tau_2\tau_3}(1+\tau_2 \tau_3)^2 \right] \\
& = & (4\pi^2\a')^5\, \prod_j \frac{\re T_j}{\tau_j}\, (1 + \tau_2 \tau_3)^2
\label{DBIdet}
\eeqa
where in the second line we have made use of (\ref{Ftermcc2}). Thus, by taking the square root of (\ref{DBIdet}) we recover the first supersymmetry condition in (\ref{SUSYD8}) for any values of $\tau_1$, $\tau_2$ and $\tau_3$. Alternatively, one can see that (\ref{Dtermcc}) is always a real quantity, and hence the D-flatness condition in (\ref{SUSYD8b}) is trivial.

How can we interpret these results? The fact that supersymmetry restricts the K\"ahler moduli is not surprising. As explained above, coisotropic D-branes are BPS because of their non-trivial worldvolume flux $F$. If we take our internal manifold to the decompactification limit, $F$ will be extremely diluted and we will approach the limit $F= 0$, where the D-brane is not BPS.\footnote{More precisely, this statement is only true for proper SU(3) holonomy manifolds like $\T^6/\IZ_2 \times \IZ_2$. This is not the case of $\T^6$, where we have a extended bulk supersymmetry and a D8-brane on (\ref{5cycle}) with $F=0$ is always BPS. The correct statement is then that such D8-brane does not preserve the same supersymmetry as a D6-brane at angles.} Hence the supersymmetry conditions of a coisotropic D-brane should constrain the K\"ahler moduli of the compactification. As will be discussed in Section \ref{effective}, this can be understood in terms of a D-brane generated superpotential.

Notice that our coisotropic D8-brane (\ref{cc}) factorizes as a product of a 1-cycle on $\T^2$ and a 4-dimensional coisotropic D-brane on $\T^4$: 
\beq
\begin{array}{rcl}
\Pi_4 & = & (\T^2)_2 \times (\T^2)_3\\
F/2\pi & = & dx^2 \wedge dx^3 - dy^2 \wedge dy^3
\end{array}
\label{cc2}
\eeq
Actually, such example of $\T^4$ coisotropic D-brane was already given in \cite{ko01}, and analyzed in great detail in \cite{az05}. There is also a {\bf K3} analogue of (\ref{cc2}), dubbed in \cite{kw06} as {\it canonical coisotropic} D-brane. An interesting fact is that the D-brane charge induced by $F$, given by its Poincar\'e dual on $\T^4$, is
\beq
[\Pi_2^F] = [(1,0)_2 \times (1,0)_3] +  [(0,1)_2 \times (0,-1)_3]
\label{PDcc2}
\eeq
This is not the homology class of any factorizable 2-cycle $(n^2,m^2) \times (n^3,m^3)$ considered in the intersecting D-brane framework, but rather a sum of two of them. These homology classes which are not factorizable as a product of 1-cycles arise in intersecting D-brane models when we consider two D-branes that have been through a recombination process \cite{bbh01,cim02b}. However, such recombined D-branes are not exact boundary states of the theory and hence it is much harder to analyze their physics (or even to prove their existence) than in the usual factorizable case. On the other hand, the coisotropic D-brane will naturally carry these non-factorizable charges, while it can still be described as an exact CFT boundary state.

In fact, all these observations made for the example (\ref{cc}) hold in general. First, any coisotropic D8-brane on $\T^6$ can be seen as a product of a 1-cycle and a coisotropic D-brane on $\T^4$. Second, the worldvolume flux $F$ on this $\T^4$ induces a D6-brane charge of the form
\beq
[\Pi_3^F]\, =\, [(n^i, m^i)_i] \otimes [\Pi_2^{jk}]
\label{D6charge}
\eeq
where $\Pi_2^{jk}$ is a Lagrangian, non-factorizable 2-cycle on $(\T^2)_j \times (\T^2)_k$. In fact, any non-factorizable 2-cycle can be decomposed as a sum of two factorizable 2-cycles \cite{rabadan01}, so we finally have that
\beq
\begin{array}{rcl}
[\Pi_3^F] & = & [(n^i, m^i)_i] \otimes [\Pi_2^\a +\Pi_2^\b]\\
& = &  [(n^i, m^i)_i] \otimes [(n^j_\a, m^j_\a)_j (n^k_\a, m^k_\a)_k + (n^j_\b, m^j_\b)_j (n^k_\b, m^k_\b)_k]
\end{array}
\label{D6charge2}
\eeq
where the intersection number $I_{\a\b}^{\T^4} = - (n^j_\a m^j_\b - n^j_\b m^j_\a)(n^k_\a m^k_\b - n^k_\b m^k_\a)$ 
is nothing but $\int_{(\T^2)_j \times (\T^2)_k} F^2/4\pi^2$, and so it cannot vanish. 
Finally, such coisotropic D8-brane is made up of a flat submanifold and a constant worldvolume flux, 
so it can be described by an exact boundary state.

An advantage of this intuitive picture is that it helps producing some new examples of coisotropic D8-branes. Indeed, we just need to choose the ten wrapping numbers above such that, for some choice of complex structure $\tau_i$, the two factorizable 3-cycles in (\ref{D6charge2}) are mutually supersymmetric, or
\beq
\begin{array}{l}
{\rm Arg}\, \left(\int_{(n^i,m^i)} dz^i\right)\, =\, \th_1\\ 
{\rm Arg}\, \left(\int_{\Pi_2^\a} dz^j \wedge dz^k\right) \, =\, {\rm Arg}\, \left(\int_{\Pi_2^\b} dz^j \wedge dz^k\right) \, =\, \th_2
\end{array}
\quad \quad \ \th_1 + \th_2 \, =\, 0\ {\rm mod}\ 2\pi
\label{dualDterm}
\eeq
where again $\{i j k\}$ is a cyclic permutation of $\{1 2 3\}$. By taking $F/2\pi$ as the Poincar\'e dual of (\ref{D6charge2}) in $\Pi_5$ we would get a coisotropic D8-brane satisfying the D-flatness condition in (\ref{SUSYD8b}), and by imposing
\beq
T_j \cdot T_k\, =\, I_{\a\b}^{\T^4}
\label{dualFterm}
\eeq
we will also satisfy the F-flatness condition.

On the other hand, the decomposition of a non-factorizable 2-cycle is not unique, and so the wrapping numbers in (\ref{D6charge2}) are not all independent. In addition, the conditions (\ref{dualDterm}) are stronger than the D-flatness condition, and one can find examples of BPS D8-branes that cannot be written in such form. It turns then more useful to describe a D8-brane directly in terms of the 5-cycle (\ref{5cycle}) and a general constant $U(1)$ bundle $F$ as
\beq
\begin{array}{rcl}
\Pi_5 & = & (n^i,m^i)_i \times (\T^2)_j \times (\T^2)_k, \\
F/2\pi & = & n^{xx}\, dx^j \wedge dx^k + n^{xy}\, dx^j \wedge dy^k + n^{yx}\, dy^j \wedge dx^k + n^{yy}\, dy^j \wedge dy^k \\
& & +\, m^{xy,\, j}\, dx^j \wedge dy^j + m^{xy,\, k}\, dx^k \wedge dy^k
\end{array}
\label{general}
\eeq
where all the $n$'s and $m$'s are integer numbers, and $\{i j k\}$ is again a cyclic permutation of $\{1 2 3\}$. It is easy to see that the F-flatness condition now reads
\beq
(T_j + i m^{xy,\, j}) \cdot (T_k + i m^{xy,\, k})\, =\, n^{xy} n^{yx} - n^{xx} n^{yy}
\label{genFterm}
\eeq
and so the only effect of $m^{xy,\, j}$, $m^{xy,\, k}$ is to shift the B-field by an integer number. For this reason we will set $m^{xy,\, j} = m^{xy,\, k} = 0$ from now on, and describe a coisotropic D8-brane $a$ in terms of the six integer numbers
\beq
\D 8_a \, :\, (n^i_a, m^i_a)_i \times (n^{xx}_a, n^{xy}_a, n^{yx}_a, n^{yy}_a)_{jk}
\label{final}
\eeq
defined by (\ref{general}). Comparing with the description (\ref{D6charge2}), we find the identifications
\beq
\begin{array}{ccc}
n^{yy}\, =\, - (n^j_\a n^k_\a + n^j_\b n^k_\b) & & n^{xx}\, =\, - (m^j_\a m^k_\a + m^j_\b m^k_\b) \\
n^{xy}\, =\, m^j_\a n^k_\a + m^j_\b n^k_\b & & n^{yx}\, =\, n^j_\a m^k_\a + n^j_\b m^k_\b
\end{array}
\label{iden}
\eeq
In the notation (\ref{final}) we also have that the pull-back of $\CF \wedge \Om$ reduces to
\beq
\CF \wedge \Om \, =\, c \cdot (n^i + i \tau_im^i)\, (n^{xx} \tau_j\tau_k + i n^{xy} \tau_j + i n^{yx} \tau_k - n^{yy})\, 
d{\rm vol}_{\Pi_5}
\label{genstable}
\eeq
where $c \in \IR$, and hence the D-flatness condition reads
\beq
m^i n^{yy} \tau_i - n^i n^{xy} \tau_j - n^i n^{yx} \tau_k\, =\,  m^i n^{xx} \tau_i \tau_j \tau_k
\label{genDterm}
\eeq
plus the additional condition $\re F \wedge \Om > 0$. As we will see below, in the case of the $\IZ_2 \times \IZ_2$ orbifold a coisotropic D8-brane can also be defined by the same set of integer numbers (\ref{final}), and the chiral spectrum as well as the effective field theory quantities of a D8-brane configuration will only depend on them.

Finally, we are interested in introducing O6-planes in our theory. For this we mod out our toroidal compactification by the orientifold action $\Om \CR (-1)^{F_L}$, where $\Om$ is the usual world-sheet parity, $F_L$ stands for left-handed spacetime fermion number, and $\CR$ is the antiholomorphic involution
\beq
\CR \,:\, (z^1, z^2, z^3)\quad \mapsto \quad (\bar{z}^1, \bar{z}^2, \bar{z}^3).
\label{R}
\eeq
In general, a magnetized D8-brane will not be left invariant under this orientifold action, so we need to include its orientifold image, 
given by\footnote{The orientifold action which takes $\D 8_a$ to $\D 8_{a'}$ may be not obvious at this stage, but is quite easy to understand from the mirror type I picture, which will be discussed in Section \ref{mirror}.}
\beq
\D 8_{a'} \, :\, (-n^i_a, m^i_a)_i \times (-n^{xx}_a, n^{xy}_a, n^{yx}_a, -n^{yy}_a)_{jk}
\label{finalprime}
\eeq
Notice that in some cases like (\ref{cc}) a D8-brane is mapped by $\Om\CR(-1)^{F_L}$ to an anti-D8-brane, so the total D8-brane charge vanishes. Nevertheless, it is easy to check that if $\D 8_a$satisfies the supersymmetry conditions (\ref{genFterm}) and (\ref{genDterm}) so will $D8_{a'}$, so both D-branes are BPS objects.

\subsection{D8-branes on $\T^6/\IZ_2 \times \IZ_2$}

The fact that we can construct coisotropic D8-branes in $\T^6$ is perhaps not surprising, given that in this compactification manifold $b_5 = 6$ and then D8-branes can wrap non-trivial 5-cycles. One may then wonder how a similar construction could be carried in general Calabi-Yau manifolds, where $b_5 = 0$ and any 5-cycle $\Pi_5$ will be trivial in homology. 

In order to get more intuition about this problem, we can consider manifolds close to $\T^6$ while carrying proper $SU(3)$ holonomy. These are toroidal orbifolds of the form $\T^6/\Gamma$, where $\Gamma \subset SO(6)$ is a discrete subgroup of $SU(3)$ but not of $SU(2)$. A simple example is given by the orbifold $\T^6/\IZ_2 \times \IZ_2$, with the $\IZ_2 \times \IZ_2$ generators acting as
\beqa
\th\, : \, (z^1, z^2, z^3) & \mapsto & (-z^1, -z^2, z^3)
\label{z2gen1}\\
\om\, : \, (z^1, z^2, z^3) & \mapsto & (z^1, -z^2, -z^3)
\label{z2gen2}
\eeqa
and where $z^i$ is the complex coordinate associated to $(\T^2)_i$. Just as in \cite{csu01}, we will consider the choice of discrete torsion such that each $\IZ_2 \times \IZ_2$ fixed point contains a collapsed 2-cycle, and so the complete orbifold cohomology is given by $(h_{11}, h_{21}) = (51,3)$.

Following the usual strategy for describing D-branes on orbifolds \cite{dm96}, we can construct a coisotropic D8-brane in $\T^6/\IZ_2 \times \IZ_2$ by considering a D8-brane on the covering space $\T^6$, then adding its images under $\IZ_2 \times \IZ_2$, and finally quotienting our theory by the orbifold action. If the D8-brane is not placed on top of any fixed point and has non-trivial Wilson lines, all the images are different and we are dealing with a {\it bulk} D8-brane whose gauge group is $U(1)$. Notice that this also applies to a D6-brane at angles, although there is one important difference. The homology class of a D6-brane $[\Pi_3]\, =\, [(n^1, m^1) \times (n^2, m^2) \times (n^3, m^3)]$ is invariant under the full orbifold action (\ref{z2gen1}), (\ref{z2gen2}), while this is not true for a D8-brane. For instance, in our $\T^6$ example (\ref{cc}), the homology class $[\Pi_5]$ is mapped into $-[\Pi_5]$ by the action of $\th$, while $F$ is mapped to $-F$. On the other hand, the action of $\om$ leaves these two quantities invariant. It is easy to see that one gets the same kind of behavior for any coisotropic D8-brane. This can be understood by the fact that a  coisotropic D8-brane on $\T^6$ will carry D8, D6 and D4-brane charges, which can be described by homology classes of 5, 3 and 1-cycles on $\T^6$. When quotienting by the $\IZ_2 \times \IZ_2$ orbifold group the non-trivial 5 and 1-cycles will be projected out, and so the total D8 and D4-brane charges (\ref{CSD8l}) should vanish. On the contrary, the induced D6-brane charge (\ref{CSD8}) will survive the orbifold projection, as can be appreciated from table \ref{project}.
\begin{table}[htdp]
\begin{center}
\begin{tabular}{|c||c|c|c|}
\hline 
 $(i=1)$ & D8 charge & D6 charge & D4 charge \\
\hline \hline
$a$ & $[\Pi_5]$ & $[\Pi_3^F]$ & $[\Pi_1^{F^2}]$ \\
$\om a$ & $[\Pi_5]$ & $[\Pi_3^F]$ & $[\Pi_1^{F^2}]$ \\
$\th a$ & $-[\Pi_5]$ & $[\Pi_3^F]$ & $-[\Pi_1^{F^2}]$ \\
$\om\th a$ & $-[\Pi_5]$ & $[\Pi_3^F]$ & $-[\Pi_1^{F^2}]$ \\
\hline
\end{tabular}
\end{center}
\caption{D-brane charges of a coisotropic D8-brane of the form (\ref{5cycle}), when embedded as a bulk D-brane on the $\IZ_2 \times \IZ_2$ orbifold. Here $[\Pi_3^F]$ is defined as the Poincar\'e dual of $[F/2\pi]$ in $\Pi_5$, while $[\Pi_1^{F^2}]$ is the Poincar\'e dual of $[F^2/4\pi^2]$. This table holds for the case where $i=1$ in (\ref{5cycle}), while in the cases $i=2$ and $i=3$, the roles of $\th$, $\om$ and $\th\om$ are interchanged.}
\label{project}
\end{table}

However, bulk coisotropic D8-branes are not very interesting from the model-building point of view. The reason is that, as we will show in the next section, the chiral spectrum of a coisotropic D8-brane only depends on the D6-brane charge that it carries. Bulk coisotropic D8-branes carry bulk D6-brane charges and, as implicit in \cite{csu01}, bulk D6-branes do only give rise to models with an even number of chiral families. Hence we would never achieve a three-family model out of bulk coisotropic D8-branes.

The way out is to consider D8-branes which are fractional with respect to the $\IZ_2 \times \IZ_2$ orbifold action, and so they can carry fractional D6-brane charge. Recall that in this particular choice of $\IZ_2 \times \IZ_2$ orbifold fractional D6-branes are $\oh$ bulk D6-branes \cite{douglas98,gomis00}, and so will be the case for fractional D8-branes. Indeed, from table \ref{project} we see that only the action of one generator of $\IZ_2 \times \IZ_2$ leaves a D8-brane invariant, while the other maps it to its anti-D8-brane. Hence, a D8-brane can only be fractional with respect to a $\IZ_2$ subgroup of $\IZ_2 \times \IZ_2$. 

\begin{figure}
\epsfysize=7cm
\begin{center}
\leavevmode 
\epsffile{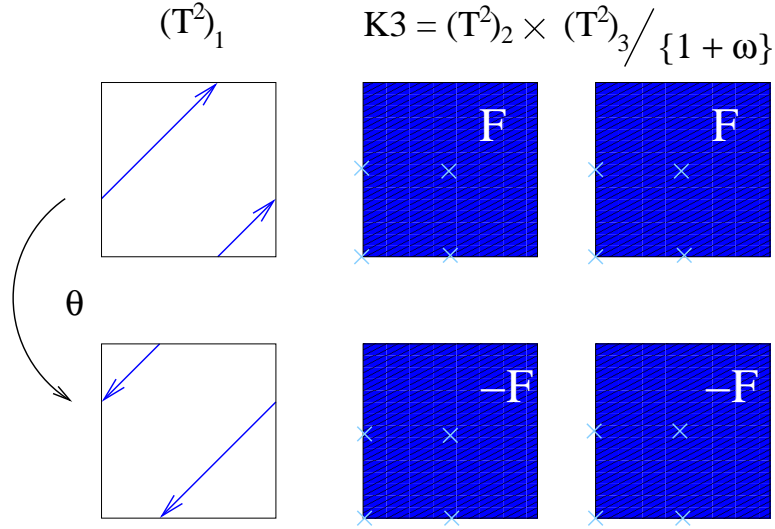}
\end{center}
\caption{Fractional coisotropic D8-brane on $\T^6/\IZ_2 \times \IZ_2$.}
\label{coiso2}
\end{figure}

For concreteness, let us consider a D8-brane of the form (\ref{5cycle}) and with $i=1$, so that it is left invariant by the action of $\om$. Generically, such D8-brane will not be on top of any $\th$-fixed point in $(\T^2)_1$, and so locally it will not know about the action of $\th$. It can thus be described by a 5-cycle of the form (see figure \ref{coiso2})
\beq
\Pi_5 \, =\, (n^1,m^1) \times \T^4/\IZ_2^\om \ \subset \ \T^2 \times \T^4/\IZ_2^\om\ \simeq \T^2 \times {\bf K3}
\label{fracD8}
\eeq
where $\IZ_2^\om$ stands for the $\IZ_2$ subgroup of $\IZ_2 \times \IZ_2$ generated by $\om$, and so $\T^2 \times {\bf K3}$ can be seen as the covering space of $\T^6/\IZ_2 \times \IZ_2$ before quotienting by $\th$. We can then describe a coisotropic D8-brane in $\T^6/\IZ_2 \times \IZ_2$ in terms of a D8-brane in $\T^2 \times {\bf K3}$ plus its image under $\th$. Moreover, because the worldvolume flux $F$ is fully contained inside ${\bf K3}$, this D8-brane can be understood as a product of a 1-cycle in $\T^2$ and a coisotropic D-brane on ${\bf K3}$.

Let us now see which D6-brane charge is carried by a fractional D8-brane. The only difference with the 
toroidal case is that $[F/2\pi]$ is now an element of $H^2({\bf K3}, \IZ)$, rather than of $H^2(\T^4, \IZ)$. 
As a result, the D6-brane charge is again of the form (\ref{D6charge}), but now $\Pi_2$ should be seen as a 
Lagrangian 2-cycle of ${\bf K3}$. As explained in \cite{bbkl02}, the difference between $H_2(\T^4, \IZ)$ 
and $H_2({\bf K3}, \IZ)$ amounts to including fractional D-branes in the orbifold $\T^4/\IZ_2$. 
More precisely, $H_2(\T^4, \IZ)$ is a sublattice of $H_2({\bf K3}, \IZ)$ which corresponds to bulk D-branes 
in $\T^4/\IZ_2$. The elements of $H_2({\bf K3}, \IZ)$ that are not in $H_2(\T^4, \IZ)$ contain the `minimal' 
2-cycles of ${\bf K3}$ and, in the orbifold limit $\T^4/\IZ_2$, they have the form
\beq
\Pi_2^{\rm fr}\, =\, \oh \Pi_2^{\rm b}\, +\, \oh \left( \sum_{i,j \in \Pi_2^{\rm b}} \epsilon_{ij} \Pi_{ij}^{\om} \right)
\label{fracK3}
\eeq
where $\Pi_2^{\rm b}$ is a bulk 2-cycle, inherited from the 2-cycles on $\T^4$ that survive the $\IZ_2$ action of $\om$. $\Pi_{ij}^\om$ stand for the $4 \times 4$ fixed points of $\T^4/\IZ_2$ or, more precisely, the 16 collapsed $\IP^1$'s that arise from such singularities. In general not all the collapsed 2-cycles contribute to (\ref{fracK3}), but only those which are crossed by $\Pi_2^{\rm b}$. The phases $\epsilon_{ij} = \pm 1$ then correspond to the two possible orientations with which the two-cycle can wrap around the blown-up $\IP^1$ labeled by $ij$. Because (\ref{fracK3}) contains the exceptional 2-cycles $\Pi_{ij}^\om$, a D-brane wrapped on it is stuck at the fixed points of $\T^4/\IZ_2$, and in fact corresponds to a $\oh$ fractional D-brane of the $\T^4/\IZ_2$ orbifold. Taking all the phases $\epsilon_{ij}$ positive corresponds to choosing one Chan-Paton phase, while taking them all negative corresponds to choosing the opposite one. The intermediate choices of $\epsilon_{ij}$'s are constrained by consistency conditions, and they can be understood as discrete Wilson lines turned on. See, e.g., \cite{bbkl02,bcms05} for more detailed discussions.

Given this description of a fractional D-brane it is now clear how to construct a coisotropic D8-brane with fractional D6-brane charge. We first consider a fractional D8-brane of the form (\ref{fracD8}) and then a pair of minimal 2-cycles  $(\Pi_2^{\rm fr})^\a$ and $(\Pi_2^{\rm fr})^\b$ such that (\ref{D6charge})-(\ref{dualDterm}) are satisfied. Just as in the toroidal case, our worldvolume flux $F$ will be given by the Poincar\'e dual of $[(\Pi_2^{\rm fr})^\a] + [(\Pi_2^{\rm fr})^\b]$ in $\Pi_5$, which means that $[F/2\pi]$ will be an element of $H^2({\bf K3}, \IZ)$ but not of the sublattice $H^2(\T^4, \IZ)$. That is, $F$ will be of the form
\beq
F^{\rm fr}\, =\, \oh F^{\rm b} + \oh F^{\rm tw} 
\label{K3flux}
\eeq
where $F^{\rm b}$ is an element of $H^2(\T^4, \IZ)$ and $F^{\rm tw}$ is the Poincar\'e dual of a sum of exceptional 2-cycles $\sum_{ij} \epsilon_{ij}\, \Pi_{ij}$ in ${\bf K3}$ which, in the orbifold limit $\T^4/\IZ_2$, can be associated to the twisted sectors of the theory. For simplicity, we will choose all the phases $\epsilon_{ij}$ to have the same sign, which amounts 
to not turning any discrete Wilson line.

The pair $(\Pi_5, F^{\rm fr})$ is a coisotropic D8-brane which is fractional with respect to $\om$, but in 
order to have a well-defined object we still need to add its image under $\th$. From the toroidal case we 
know how $\Pi_5$ and $F^{\rm b}$ transform under the action of $\th$. The behavior of $F^{\rm tw}$ is more 
subtle and depends on the choice of discrete torsion. Recall that in a $\IZ_2 \times \IZ_2$ orbifold the 
choice of discrete torsion amounts to specifying the action of $\th$ on the collapsed 
$\IP^1$'s of ${\bf K3} \simeq \T^4/\IZ_2^\om$. The two possibilities are
\beq
\th\, : \, \IP^1\, \mapsto\, \eta \IP^1
\label{dtorsion}
\eeq
with $\eta = \pm 1$. In our case $\eta =1$, i.e., the collapsed two cycles do not change orientation under the action of $\th$, and so the phases $\eps_{ij}$ in (\ref{fracK3}) do not change when we consider the image of $\Pi_2^F$ under $\th$. We similarly find that the action of $\th$ on a fractional coisotropic D8-brane is given by
\beq
\begin{array}{rcl}
\, [\Pi_5] & \stackrel{\th}{\mapsto} & - [\Pi_5] \\
\, \oh F^{\rm b} +\oh F^{\rm tw}& \stackrel{\th}{\mapsto} &  - \oh F^{\rm b} + \frac{\eta}{2} F^{\rm tw}
\end{array}
\label{dtorsion2}
\eeq
where we have chosen $F^{\rm b}$ as in (\ref{final}). The choice $\eta = 1$ then implies that, while the untwisted D6-brane charge is left invariant under the action of $\th$, the twisted D6-brane charge is flipped by a minus sign, and so the total twisted D6-brane charge vanishes (see table \ref{project2}). This is indeed what we would expect, since for this choice of discrete torsion the RR twisted fields coupling to an A-brane are projected out, and hence such D-brane should not carry any twisted charge.
\begin{table}[htdp]
\begin{center}
\begin{tabular}{|c||c|c|c|c|}
\hline 
 $(i=1)$ & D8 charge & D6 charge & tw. D6 charge & D4 charge \\
\hline \hline
$a$ & \hspace*{.2cm} $[\Pi_5]$ &  $[\Pi_3^{F^{\rm b}}]$ &  \hspace*{.2cm} $[\Pi_3^{F^{\rm tw}}]$ &  \hspace*{.2cm} $[\Pi_1^{F^2}]$ \\
$\th a$ & $-[\Pi_5]$ & $[\Pi_3^{F^{\rm b}}]$ & $-[\Pi_3^{F^{\rm tw}}]$ & $-[\Pi_1^{F^2}]$ \\
\hline \hline
$\tilde{a}$ &  \hspace*{.2cm} $[\Pi_5]$ & $[\Pi_3^{F^{\rm b}}]$ & $-[\Pi_3^{F^{\rm tw}}]$ &  \hspace*{.2cm} $[\Pi_1^{F^2}]$ \\
$\th \tilde{a}$ & $-[\Pi_5]$ &  $[\Pi_3^{F^{\rm b}}]$ &  \hspace*{.2cm} $[\Pi_3^{F^{\rm tw}}]$ &  $-[\Pi_1^{F^2}]$ \\
\hline
\end{tabular}
\end{center}
\caption{D-brane charges of a fractional coisotropic D8-brane and its image under $\th$, given the choice of discrete torsion $\eta =1$. The upper two rows represent a $\IZ_2^\om$-fractional D8-brane $a$ with a definite choice of Chan-Paton phase (say $\a = +1$), while the two lower rows correspond to the opposite choice ($\a = - 1$). These four D8-branes are then the constituents of a bulk coisotropic D8-brane.}
\label{project2}
\end{table}

We can now understand a bulk D8-brane in term of its fractional components, as shown in table \ref{project2}. Indeed, in order to construct a bulk coisotropic D8-brane we need to consider a fractional D8-brane $a$ with some choice of Chan-Paton phase $\a$ and a second D8-brane $\tilde{a}$ with the opposite choice of $\a$. We then add their images under $\th$ and obtain a coisotropic D8-brane in the regular representation of the $\IZ_2 \times \IZ_2$ orbifold group. Notice that each of the pairs $(a, \th a)$ and $(\tilde{a}, \th\tilde{a})$ has vanishing D8, D4 and twisted D6-brane charge. In addition, each pair generates a $U(1)$ vector multiplet, and there are no gauge bosons arising from open strings in mixed sectors like $a\tilde{a}$. Hence the total gauge group of a coisotropic D8-brane in a regular representation is given by $U(1) \times U(1)$. In particular, this means that when a bulk D8-brane approaches an orbifold singularity its gauge group gets enhanced as
\beq
U(1)\ \longrightarrow \ U(1) \times U(1)
\label{enhancement}
\eeq 
in contrast to the enhancement $U(1) \rightarrow U(2)$ that occurs for D6-branes \cite{douglas98}. As seen above, each of the $U(1)$ factors in (\ref{enhancement}) can be seen as an independent D8-brane whose net RR charges are those of a fractional D6-brane. More precisely, the net RR charge of a pair $(a, \th a)$ is given by $[\Pi_3^{F^{\rm b}}]$, which is specified by the six integer numbers (\ref{final}) of the previous $\T^6$ case. These fractional D8-branes will be the building blocks that we will consider in order to construct $\CN=1$ chiral models.

In fact, $F^{\rm b}$ will be the only quantity to consider when constructing a fractional BPS D8-brane in $\T^6/\IZ_2 \times \IZ_2$, rather than the full $F^{\rm fr}$. This can be seen from the fact that $F^{\rm b}$ is the only worldvolume flux present for a generic bulk D8-brane. Taking such bulk D8-brane to an orbifold singularity and splitting it as in table \ref{project2} should not affect its BPSness. Thus, if a bulk D8-brane is BPS for some choice of closed string moduli so will be its fractional components. These fractional components inherit the same $F^{\rm b}$ as the bulk D8-brane, but now have a new contribution to $F$ coming from $F^{\rm tw}$. By construction, the latter should not be involved in the supersymmetry equations (\ref{SUSYD8}).

A more direct way to reach the same conclusion is to follow the standard procedure to construct D-branes on orbifolds \cite{dm96}. According to such prescription, the BPSness of a fractional D-brane on $\T^6/\IZ_2 \times \IZ_2$ is specified by its BPSness in the covering space $\T^6$, with the only condition that the supersymmetry preserved by such D-brane on $\T^6$ needs to be compatible with the bulk supersymmetry preserved by the $\IZ_2 \times \IZ_2$ action. Again, if a magnetized D8-brane is BPS or not in $\T^6$ only depends on $\Pi_5$ and in $F^{\rm b}$, and so the precise form of $F^{\rm tw}$ will not matter when constructing a BPS D8-brane.\footnote{In the usual orbifold language, the twisted charges $F^{\rm tw}$ do not appear as such, but are encoded in the matrices $\g_{\th}$, $\g_{\om}$ acting on the Chan-Paton degrees of freedom. In this paper we are using a geometric approach which does not make use of the explicit form of $\g_{\th}$ and $\g_{\om}$, but it would be interesting to rederive our results from this more standard procedure.}

To summarize, we should only concentrate on $\Pi_5$ and $F^{\rm b}$ when applying the D-flatness and F-flatness conditions for a fractional coisotropic D8-brane. Because both quantities are specified by the toroidal quanta (\ref{final}), we will obtain similar constraints on the untwisted K\"ahler and complex structure moduli as those found for the $\T^6$ case. In fact we will see that, for most purposes, we can treat a fractional D8-brane on $\T^6/\IZ_2 \times \IZ_2$ as a D8-brane on $\T^6$.

Similarly, $\Pi_5$ and $F^{\rm b}$ will be the only quantities to consider when checking the consistency of a coisotropic D8-brane model. More precisely, in the present $\T^6/\IZ_2 \times \IZ_2$ orbifold background the zero modes of the RR $C_9$ and $C_5$ forms are projected out, and a coisotropic D8-brane may only contribute to $C_7$ RR divergences via its induced untwisted D6-brane charge $[\Pi_3^{F^{\rm b}}]$. If in addition we mod out our theory by the orientifold action $\Om \CR (-1)^{F_L}$ specified by (\ref{R}), the only surviving components of $C_7$ will be those of the form $(C_7)_{x^1 x^2 x^3}$ and $(C_7)_{x^i y^j y^k}$, yielding four independent RR tadpole conditions.

In order to write down such conditions, let us consider a configuration made up of fractional D6 and coisotropic D8-branes, with topological data
\beqa
\D 6_i  & :  & N_i \ (n_i^1, m_i^1)_1 \times  (n_i^2, m_i^2)_2 \times  (n_i^3, m_i^3)_3 \nonumber \\
\D 8_a & : & N_a \ (n_a^1, m_a^1)_1 \times  (n_a^{xx}, n_a^{xy}, n_a^{yx}, n_a^{yy})_{23}
\nonumber \\
\D 8_b & : & N_b \ (n_b^2, m_b^2)_2  \times  (n_b^{xx}, n_b^{xy}, n_b^{yx}, n_b^{yy})_{31}
\label{stackdata} \\
\D 8_c & : & N_c \ (n^3_c, m^3_c)_3  \times  (n_c^{xx}, n_c^{xy}, n_c^{yx}, n_c^{yy})_{12} 
\nonumber
\eeqa
where, following the conventions in \cite{csu01}, the number of D-branes is normalized such that $N_\a$ is an even number yielding a $U(N_\a/2)$ gauge group. Cancellation of RR tadpoles then reads
\beqa
\sum_{i,a,b,c} N_i n_i^1 n_i^2 n_i^3 - N_a n_a^1 n_a^{yy} - N_b n_b^2 n_b^{yy} - N_c n_c^3 n_c^{yy}
& = & 16 \nonumber \\[0.2cm]
\sum_{i,a,b,c} N_i n_i^1 m_i^2 m_i^3
- N_a n_a^1 n_a^{xx} + N_b m_b^2 n_b^{xy} + N_c m_c^3 n_c^{yx} 
& = & -16 \nonumber \\[0.2cm]
\sum_{i,a,b,c} N_i m_i^1 n_i^2 m_i^3
+ N_a m_a^1 n_a^{yx} -  N_b n_b^2 n_b^{xx} + N_c m_c^3 n_c^{xy}
& = & -16
\label{tadpoled6} \\[0.2cm]
\sum_{i,a,b,c} N_i m_i^1 m_i^2 n_i^3
+ N_a m_a^1 n_a^{xy} + N_b m_b^2 n_b^{yx} - N_c n_c^3 n_c^{xx} 
& = & -16
\nonumber
\eeqa
where we have also included the contribution of the O6-planes introduced by the orientifold quotient.

In addition, in this class of orientifold backgrounds there may be extra consistency conditions which are invisible to one-loop divergences in the worldsheet. Such extra constraints take into account the cancellation of those D-brane charges which cannot be detected by supergravity, but only by a K-theory computation or by using D-brane probes \cite{uranga00}. For the $\IZ_2 \times \IZ_2$ orientifold at hand and for D6-branes at angles, these extra constraints were computed in \cite{ms04}. One can repeat the same D-brane probe argument to include D8-branes, obtaining
\beqa
\sum_{i,a,b,c} N_i m_i^1 m_i^2 m_i^3 - N_a m_a^1 n_a^{xx} - N_b m_b^2 n_b^{xx} - N_c m_c^3 n_c^{xx}
& = & 0\ {\rm mod}\ 4 \nonumber \\[0.2cm]
\sum_{i,a,b,c} N_i m_i^1 n_i^2 n_i^3
- N_a m_a^1 n_a^{yy} + N_b n_b^2 n_b^{yx} + N_c n_c^3 n_c^{xy} 
& = & 0\ {\rm mod}\ 4 \nonumber \\[0.2cm]
\sum_{i,a,b,c} N_i n_i^1 m_i^2 n_i^3
+ N_a n_a^1 n_a^{xy} -  N_b m_b^2 n_b^{yy} + N_c n_c^3 n_c^{yx}
& = & 0\ {\rm mod}\ 4
\label{Ktheoryd6} \\[0.2cm]
\sum_{i,a,b,c} N_i n_i^1 n_i^2 m_i^3
+ N_a n_a^1 n_a^{yx} + N_b n_b^2 n_b^{xy} - N_c m_c^3 n_c^{y} 
& = & 0\ {\rm mod}\ 4
\nonumber
\eeqa
which imply the cancellation of K-theoretical $\IZ_2$ charges that both D6 and D8-branes may carry. Finally, the CFT analysis of \cite{mss06} suggests that certain coisotropic D8-branes could carry additional $\IZ_2$ charges, again invisible to supergravity. As stated in \cite{mss06}, it is not clear if such additional $\IZ_2$ charges could give rise to extra consistency constraints or if (\ref{tadpoled6}) and (\ref{Ktheoryd6}) gather all the consistency conditions of our compactification. For simplicity, we will consider the latter to be the case and leave the analysis of the full set of torsional K-theory charges for future work.

\section{Coisotropic branes and chirality \label{chirality}}

One interesting property of coisotropic D8-branes is that, just like intersecting D6-branes, they give rise to chiral fermions upon compactification. Recall that, in the intersecting D6-brane framework, chiral fermions arise at the intersection locus of the 3-cycles $\Pi_3^{D6_a}$ and $\Pi_3^{D6_b}$ that two D6-branes wrap. These intersections are points in the internal manifold $\CM_6$, and they naturally give rise to $D=4$ fermions in the effective theory \cite{bdl96}. The multiplicity of chiral fermions is given by the signed number of intersections
\beq
I_{ab}\, =\, [\Pi_3^{D6_a}] \circ [\Pi_3^{D6_b}]
\label{inter}
\eeq
which only depends on the D6-brane homological charges $[\Pi_3^{D6_a}]$ and $[\Pi_3^{D6_b}]$.

If we now consider two D8-branes $\D 8_a$ and $\D 8_b$ they may also intersect, but they would do so in a 4-cycle $\CS_4$ and there is a priori no reason to expect $D=4$ chirality to arise from such intersection. However, because coisotropic D8-branes carry a worldvolume flux $F$, 
there will be a relative flux $F_{ab} = F_b - F_a$ coupling to the open strings in the $\D 8_a \D 8_b$ sector. This flux will modify the internal Dirac index of the $D=8$ theory compactified on $\CS_4$ and, in particular, it will be a source of $D=4$ chirality via the coupling $\int_{\CS_4} F_{ab}^2$. 
The same story will apply to a $\D 6_a$-$\D 8_b$ system. Now the intersection locus is given by a 2-cycle $\CS_2$, and a natural source of $D=4$ chirality will be given by the quantity $\int_{\CS_2} F_{ab}$. 

We thus see that, when including D8-branes into the picture, the source of chirality in type IIA compactifications is not only given by the intersection of D-branes, but rather by a combination of intersection and magnetization mechanism. This reminds of the case of type IIB theory, where such combination also arises from compactification with D5 or D7-branes. From the type IIB literature we know that computing the chiral spectrum in those cases may be quite involved, and that it requires describing our D-branes as coherent sheaves and computing the so-called `Ext' groups among them \cite{sheaves}. In principle we could apply a similar procedure to compute the matter content of our $\D 8_a$-$\D 8_b$ and $\D 6_a$-$\D 8_b$ system. However, we will see below that for coisotropic D8-branes on $\T^6/\IZ_2 \times \IZ_2$ the number of chiral fermions can be expressed in a rather simple way. Indeed, let $[\Pi_3^{D8}]$ be the D6-brane charge induced on a coisotropic D8-brane. Then the net chirality on a $\D 8_a$-$\D 8_b$ and $\D 6_a$-$\D 8_b$ system is given by
\beq
I_{ab}\, =\, [\Pi_3^{D8_a}] \circ [\Pi_3^{D8_b}]\quad {\rm and} \quad
I_{ab}\, =\, [\Pi_3^{D6_a}] \circ [\Pi_3^{D8_b}]
\label{inter2}
\eeq
respectively. In fact, we would expect such expressions to be valid for general Calabi-Yau manifold $\CM_6$. Notice that when $[\Pi_3^{D8}]$ is trivial in the homology of $\CM_6$ then any of these intersection numbers will vanish, but then our D8-brane could never be a BPS object. Hence the source of stability for a coisotropic D8-brane (i.e., the worldvolume flux $F$) turns out to be also the source of $D=4$ chirality.

\subsection{D6-D6 chirality}

Let us first review the case where chirality arises from two intersecting D6-branes. The simplest case is that of $\T^6 = \T^2 \times \T^2 \times \T^2$ and two D6-branes at angles
\beq
\begin{array}{rcl}
\D 6_a  & :  & N_a \ (n_a^1, m_a^1)_1 \times  (n_a^2, m_a^2)_2 \times  (n_a^3, m_a^3)_3 \\
\D 6_b  & :  & N_b \ (n_b^1, m_b^1)_1 \times  (n_b^2, m_b^2)_2 \times  (n_b^3, m_b^3)_3 
\end{array}
\label{D6D6i}
\eeq
yielding a gauge group $U(N_a) \times U(N_b)$. The number of chiral fermions in the bifundamental representation $(N_a, \ov{N_b})$ is given by the intersection number (\ref{inter}), which is computed by taking the Poincar\'e dual 3-forms
\beq
\begin{array}{rcl}
\om_a  & =  & - (n_a^1 dy^1 - m_a^1 dx^1) \wedge (n_a^2 dy^2 - m_a^2 dx^2) \wedge (n_a^3 dy^3 - m_a^3 dx^3) \\
\om_b  & =  & - (n_b^1 dy^1 - m_b^1 dx^1) \wedge (n_b^2 dy^2 - m_b^2 dx^2) \wedge (n_b^3 dy^3 - m_b^3 dx^3)
\end{array}
\label{D6D6ii}
\eeq
and integrating them over $\T^6$ to yield
\beq
I_{ab} \, =\, \int_{\T^6} \om_a \wedge \om_b, =\, (n_a^1m_b^1 - n_b^1m_a^1) \cdot (n_a^2m_b^2 - n_b^2m_a^2) \cdot (n_a^3m_b^3 - n_b^3m_a^3) 
\label{D6D6iii}
\eeq

\begin{figure}
\epsfysize=4.8cm
\begin{center}
\leavevmode 
\epsffile{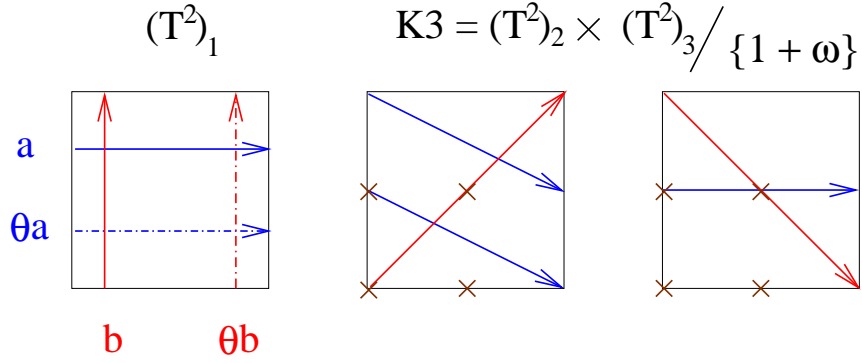}
\end{center}
\caption{Chirality from a D6-D6 system on $\T^6/\IZ_2 \times \IZ_2$.}
\label{D6D6orb}
\end{figure}

When considering the orbifold $\T^6/\IZ_2 \times \IZ_2$ the situation is more involved but, as shown in \cite{csu01} by means of CFT arguments, the number of chiral fermions is again given by (\ref{D6D6iii}), and the only difference is that the gauge group is now $U(N_a/2) \times U(N_b/2)$. Let us rederive this result by the more geometric approach followed in \cite{bbkl02}. Since $N_a$ is an even number, we can separate a stack of $N_a/2$ D6$_a$-branes away from the fixed points on $(\T^2)_1$, the remaining $N_a/2$ D6$_a$-branes being their images under $\th$. We can repeat the same procedure with $N_b/2$ D6$_b$-branes, without changing the gauge group or chiral spectrum of our configuration \cite{clll04} (see figure \ref{D6D6orb}). Just like the fractional D8-brane discussed above, these D6-branes see the geometry $(\T^2)_1 \times \T^4/\IZ_2^\om$, so they wrap a 3-cycle of the form
\beq
\begin{array}{rcl}
\D 6_\a  & :  & N_\a/2 \ (n_\a^1, m_\a^1)_1 \times  \left( \oh \Pi_{2, \a}^{\rm b} + \oh \Pi_{2, \a}^{\rm tw} \right)\\ \vspace*{.2cm}
\D 6_{\th \a}  & : &  N_\a/2 \ (n_\a^1, m_\a^1)_1 \times  \left( \oh \Pi_{2, \a}^{\rm b} - \oh \Pi_{2, \a}^{\rm tw} \right) \\
& & \Pi_{2, \a}^{\rm b}\, =\, (n_\a^2, m_\a^2)_2 \times  (n_\a^3, m_\a^3)_3
\end{array}
\label{D6D6orbi}
\eeq
where $\a =a, b$ and we have decomposed a minimal 2-cycle of $\T^4/\IZ_2$ as in (\ref{fracK3}). Again, the relative sign in $\Pi_{2, \a}^{\rm tw}$ follows from the specific choice of discrete torsion $\eta = 1$ in (\ref{dtorsion}).

In order to compute the chiral spectrum we sum over the intersections of $\D 6_a \D 6_b$ and $\D 6_a \D 6_{\th b}$, the other two sectors being mapped to these by the action of $\th$. We then find
\beq
\begin{array}{rcl}
I_{ab}& = & [\Pi_3^{D6_a}] \circ [\Pi_3^{D6_b}] + [\Pi_3^{D6_a}] \circ [\Pi_3^{D6_{\th b}}]\\
& = & \left([(n_a^1, m_a^1)] \circ_{\T^2} [(n_b^1, m_b^1)]\right) \cdot \left(\oh [\Pi_{2, a}^{\rm b}] \circ_{\bf K3} [\Pi_{2, b}^{\rm b}]\right)\\
& = & \left(-(n_a^1m_b^1 - n_b^1m_a^1)  \right) \cdot \left(- (n_a^2m_b^2 - n_b^2m_a^2) \cdot (n_a^3m_b^3 - n_b^3m_a^3) \right)
\end{array}
\label{D6D6orbii}
\eeq
where we have used the fact that $[\Pi_{2, \a}^{\rm b}] \circ_{\bf K3} [\Pi_{2, \a}^{\rm tw}] = 0$ and that the embedding of $H_2(\T^4, \IZ)$ 
into $H_2({\bf K3}, \IZ)$ involves a factor of two \cite{bbkl02}.\footnote{Recall that in our conventions  $\int_{(\T^2)_i}  dy^i \wedge dx^i = 1$ (see footnote \ref{conv}), and so there is a relative sign with respect to the usual intersection product used for $\T^2$.} Finally, one can check that both summands in the first line of (\ref{D6D6orbii}) have the same sign, and so there is no vector-like pairs arising from the D6-D6 system.

\subsection{D6-D8 chirality}

\begin{figure}
\epsfysize=4cm
\begin{center}
\leavevmode 
\epsffile{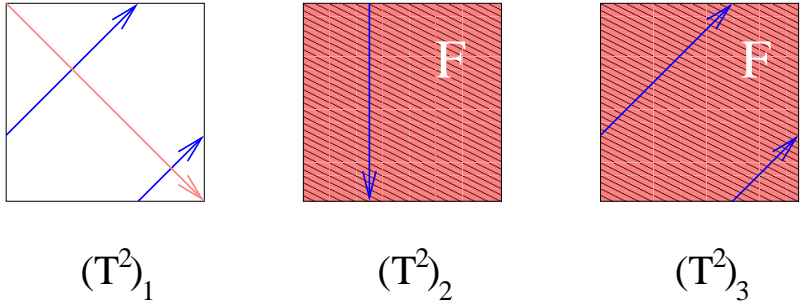}
\end{center}
\caption{Chirality from a D6-D8 system on $\T^6$.}
\label{D6D8}
\end{figure}

Let us now consider a D6$_a$-D8$_b$ system on $\T^6$. In particular, let us take the case where the direction not filled by the D8-brane is on $(\T^2)_i$, so that we have
\beq
\begin{array}{rcl}
\D 6_a  & :  & N_a \ (n_a^1, m_a^1)_1 \times  (n_a^2, m_a^2)_2 \times  (n_a^3, m_a^3)_3 \\
\D 8_b  & :  & N_b \ (n_b^i, m_b^i)_i \times  (n_b^{xx}, n_b^{xy}, n_b^{yx}, n_b^{yy})_{jk}
\end{array}
\label{D6D8i}
\eeq
as illustrated in figure \ref{D6D8}, with $i=1$. Both D-branes intersect $|n_a^i m_b^i - n_b^i m_a^i|$ times on $(\T^2)_i$, and  
each intersection is a $\T^2$ of the form $\Pi_{2}^{D6_a}\, =\, \{ (n_a^j, m_a^j)_j \times  (n_a^k, m_a^k)_k\}$. Thus, each intersection 
can be seen as a $D=6$ theory compactified on $\T^2$, with a $D=6$ chiral fermion arising from the $\D 6_a \D 8_b$ sector. Such $D=6$ theory is deformed by the presence of a relative flux $F_{ab}$, which is nothing but the pull-back of $F^{D8_b}$ on $\Pi_{2}^{D6_a}$. The multiplicity of chiral fermions associated to the Landau levels of this toron system is given by the index
\beq
\int_{\Pi_{2}^{D6_a}} F_b\, =\, \int_{\T^4} \tilde{\om}_a \wedge F_b \, =\, [\Pi_2^{D6_a}] \circ_{\T^4} [\Pi_2^{D8_b}] \, =\, I_{ab}^{\T^4}
\label{D6D8ii}
\eeq
where $\tilde{\om}_a$ is the Poincar\'e dual of $\Pi_{2}^{D6_a}$ on $\T^4$. If we multiply this spectrum by the number of intersections on $(\T^2)_i$ we arrive to the total multiplicity
\beq
I_{ab} \, = \, I_{ab}^{(\T^2)_i} \cdot I_{ab}^{(\T^4)_{jk}}\, =\, [\Pi_3^{D6_a}] \circ [\Pi_3^{D8_b}]
\label{D6D8iii}
\eeq
as advanced in (\ref{inter2}). In terms of the integer numbers (\ref{D6D8i}) we have a total of
\beq
I_{ab} \, = \, - (n_a^im_b^i - n_b^im_a^i) \cdot (n_a^jn_a^kn^{xx}_b +n_a^jm_a^kn^{xy}_b + m_a^jn_a^kn^{yx}_b + m_a^jm_a^kn^{yy}_b)
\label{D6D8iv}
\eeq
chiral multiplets transforming in the bifundamental representation $(N_a, \ov{N_b})$. 

In order to compute the chiral spectrum of the D6$_a$-D8$_b$ system in the  $\T^6/\IZ_2 \times \IZ_2$ 
orbifold we can follow the same strategy used for intersecting D6-branes. We separate both D6 and 
D8-branes away from the $\IZ_2 \times \IZ_2$ fixed points on one of the two-tori, say $(\T^2)_1$, 
and then we describe them as fractional branes on $\T^2 \times \T^4/\IZ_2$. The main difference with 
the computation of $I_{ab}^{\T^4}$ is that now $\Pi_2^{D6_a}$ and $F_b$ have each a component from 
the twisted sectors of ${\bf K3}$. However, the twisted contributions to $I_{ab}^{\bf K3}$ will 
cancel once that we have summed over the sectors $\D 8_b$ and $\D 8_{\th b}$, and so we will recover (\ref{D6D8iv}). 
The only difference with $\T^6$ will be that now the gauge group is given by $U(N_a/2) \times U(N_b/2)$, 
and so (\ref{D6D8iv}) indicates the number of chiral multiplets in the representation $(N_a/2, \ov{N_b}/2)$.

\subsection{D8-D8 chirality}

Finally, let us consider a D8$_a$-D8$_b$ system. On $\T^6$ our coisotropic D8-branes are described by
\beq
\begin{array}{rcl}
\D 8_a  & :  & N_a \ (n_a^i, m_a^i)_i \times  (n_a^{xx}, n_a^{xy}, n_a^{yx}, n_a^{yy})_{jk} \\
\D 8_b  & :  & N_b \ (n_b^{i'}, m_b^{i'})_{i'} \times  (n_b^{xx}, n_b^{xy}, n_b^{yx}, n_b^{yy})_{j'k'}
\end{array}
\label{D8D8i}
\eeq
where $\{ijk\}$ and $\{i'j'k'\}$ are both even permutations of $\{123\}$. Although D8$_a$ and D8$_b$ will always intersect on a $\T^4$, we can distinguish between two different cases:

\begin{figure}
\epsfysize=7.5cm
\begin{center}
\leavevmode 
\epsffile{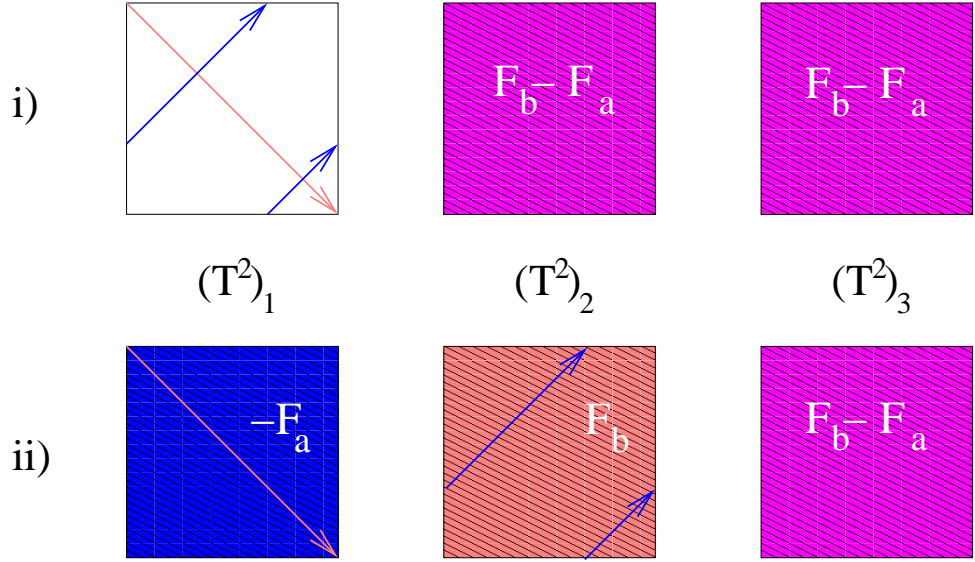}
\end{center}
\caption{Chirality from a D8-D8 system on $\T^6$.}
\label{D8D8}
\end{figure}

\begin{itemize}

\item{$i = i'$ [figure \ref{D8D8}, $i)$]}

In this case D8$_a$ and D8$_b$ intersect $|n_a^i m_b^i - n_b^i m_a^i|$ on $(\T^2)_i$, and they overlap on $(\T^2)_j \times (\T^2)_k$. On each intersection we have a $D=8$ gauge theory compactified on $\T^4$, with a relative flux $F_{ab} = F_b - F_a$ acting on the D8$_a$D8$_b$ sector. The multiplicity associated to the Landau levels of this compactification is now given by
\beq
- \oh \int_{(\T^2)_j \times (\T^2)_k} \!\!\!\!\!\!\!\!\!\!\!\!\!\!\!\!\! (F_b - F_a)^2 \, =\, (n_b^{xx}-n_a^{xx})(n_b^{yy}- n_a^{yy}) - (n_b^{xy} - n_a^{xy}) (n_b^{yx} - n_a^{yx})
\label{D8D8ii}
\eeq
and so the total number of chiral fermions is 
\beq
- (n_a^i m_b^i - n_b^i m_a^i) \cdot \left( (n_b^{xx}-n_a^{xx})(n_b^{yy}- n_a^{yy}) - (n_b^{xy} - n_a^{xy}) (n_b^{yx} - n_a^{yx}) \right)
\label{D8D8iii}
\eeq

\item{$i \not= i'$ [figure \ref{D8D8}, $ii)$]}

In this case both D8-branes intersect just once and the $\T^4$ where they overlap is of the form $(\T^4)_{ab} = (n_a^i,m_a^i) \times (n_b^{i'}, m_b^{i'}) \times (\T^2)_k$, with $i \not= k \not= i'$. The multiplicity of the D8$_a$D8$_b$ is now given by the integral
\beq
\begin{array}{rcl}
- \int_{(\T^4)_{ab}} \oh (F_b - F_a)^2 & = & \int_{(n_a^i,m_a^i) \times (n_b^{i'}, m_b^{i'}) \times (\T^2)_k} F_a \wedge F_b\\
& = & \int_{\T^6} (n_a^i dy^i - m_a^i dx^i) \wedge F_a \wedge  (n_b^{i'} dy^{i'} - m_b^{i'} dx^{i'}) \wedge F_b \\
& = & \int_{\T^6} \om_a \wedge \om_b \, =\, [\Pi_3^{D8_a}] \circ [\Pi_3^{D8_b}]
\end{array}
\label{D8D8iv}
\eeq
where $[\om_\a]$ is the Poincar\'e dual of $[\Pi_3^{D8_\a}]$. In terms of the D8-brane quanta (\ref{D8D8i}) we have, for $i' = j$,
\beq
\begin{array}{rcl}
I_{ab} & = & n_a^i m_b^{i'} (n_a^{yy}  n_b^{xx} -  n_a^{yx} n_b^{yx})  - m_a^{i}  n_b^{i'} (n_a^{xx} n_b^{yy} - n_a^{xy} n_b^{xy})\\
& &  - n_a^i n_b^{i'} (n_a^{xx}  n_b^{yx} -  n_a^{xy} n_b^{xx}) - m_a^{i}  m_b^{i'} (n_a^{yy} n_b^{xy} - n_a^{yx} n_b^{yy})
\end{array}
\label{D8D8v}
\eeq
and for $i' = k$
%
%
we just need to interchange $(i \leftrightarrow i')$ and $(a \leftrightarrow b)$ in (\ref{D8D8v}) in order to compute $-I_{ab}$.

\end{itemize}

Notice that (\ref{D8D8v}) can be put in the form (\ref{inter2}), but that this is not the case for (\ref{D8D8iii}). This is because $\T^6$ contains non-trivial 5 and 1-cycles, and so there are non-trivial RR charges associated to both D8 and D4-branes wrapped on such cycles. One would expect such charges to play a role when computing the chiral spectrum of a compactification, in the same sense that a D4-D8 system gives rise to $D=4$ chirality when both D-branes have a non-trivial intersection. In Calabi-Yau manifolds with $b_1 = 0$ the only conserved charges should be related to 3-cycles $\Pi_3$ and then we should be able to compute the chiral spectrum of a D8-D8 system by means of (\ref{inter2}).

\begin{figure}
\epsfysize=8cm
\begin{center}
\leavevmode 
\epsffile{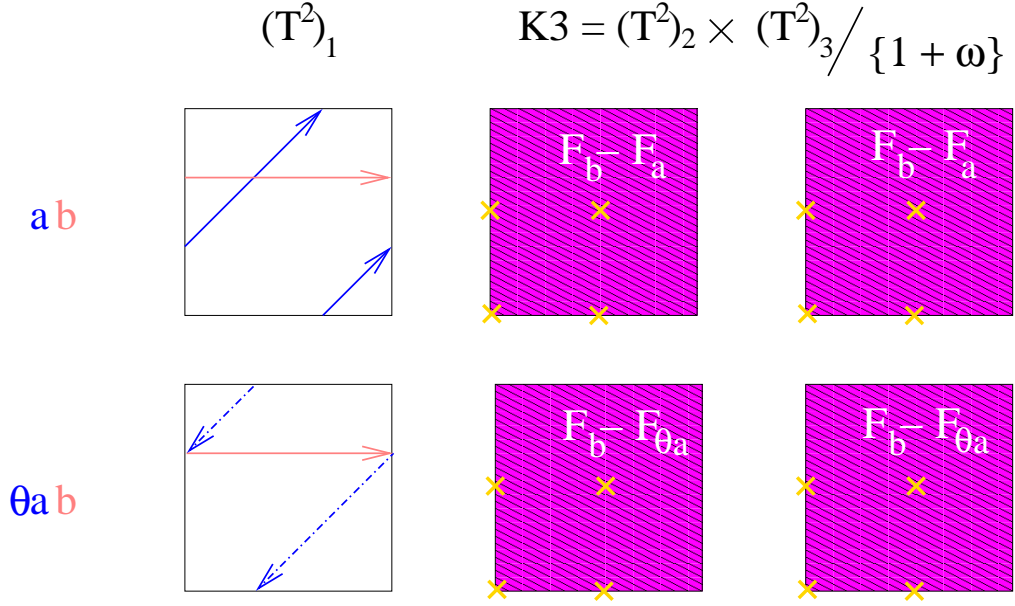}
\end{center}
\caption{Chirality from a D8-D8 system on $\T^6/\IZ_2 \times \IZ_2$.}
\label{D8D8orb}
\end{figure}

This is indeed the case of $\T^6/ \IZ_2 \times \IZ_2$, which we now analyze. We again describe our fractional D8-branes in terms of the covering space $\T^2 \times \T^4/\IZ_2$, as shown in figure \ref{coiso2}. Recall that now $F_a$ and $F_b$ are 2-forms in 
{\bf K3} and that they are mapped to (\ref{dtorsion2}) by the action of $\th$, so that we have
\beq
\begin{array}{rcl}
\D 8_\a  & :  & N_\a/2 \ (n_\a^i, m_\a^i)_i \times  {\bf K3}, \quad F_\a^{\rm fr}\, =\, \oh F_\a^{\rm b} + \oh F_\a^{\rm tw} \\ \vspace*{.2cm}
\D 8_{\th \a}  & : &  N_\a/2 \ (-n_\a^i, -m_\a^i)_i \times  {\bf K3}, \quad F_{\th\a}^{\rm fr}\, =\, -\oh F_\a^{\rm b} + \oh F_\a^{\rm tw} \\
F_\a^{\rm b} & = & n^{xx}_\a\, dx^j \wedge dx^k + n^{xy}_\a\, dx^j \wedge dy^k + n^{yx}_\a\, dy^j \wedge dx^k + n^{yy}_\a\, dy^j \wedge dy^k
\end{array}
\label{D8D8orbi}
\eeq
Adding up the contributions from the D8$_a$-D8$_b$ and  D8$_{\th a}$-D8$_b$ sectors of figure \ref{D8D8orb}  we arrive to a chiral spectrum given by
\beq
\begin{array}{rcl}
I_{ab} & = & (n_a^i m_b^i - n_b^i m_a^i) \cdot \left(\int_{\bf K3} \oh (F_b^{\rm fr} - F_{a}^{\rm fr})^2 - \int_{\bf K3} \oh (F_b^{\rm fr} - F_{\th a}^{\rm fr})^2 \right) \\
& = & - (n_a^i m_b^i - n_b^i m_a^i) \cdot \int_{\bf K3} F_b^{\rm fr} \wedge \left( F_{a}^{\rm fr} - F_{\th a}^{\rm fr} \right)\\
 & = & - (n_a^i m_b^i - n_b^i m_a^i) \cdot \oh \int_{\bf K3} F_b^{\rm b} \wedge  F_a^{\rm b} \, =\, [\Pi_3^{D8_a}] \circ [\Pi_3^{D8_b}]
\end{array}
\label{D8D8orbii}
\eeq
which can indeed be expressed as the intersection product of two 3-cycles. As usual, this quantity only depends on the bulk wrapping and magnetic numbers (\ref{D8D8i}), in terms of which it reads
\beq
I_{ab}\, =\, (n_a^i m_b^i - n_b^i m_a^i) \cdot (n_a^{xx}n_b^{yy} + n_a^{yy}n_b^{xx} - n_a^{xy}n_b^{yx} - n_a^{yx}n_b^{xy}) 
\label{D8D8orbiii}
\eeq

Finally, the same procedure can be applied to the D8-D8 system of figure \ref{D8D8}, $ii)$ in order to compute its chiral spectrum when embedded in $\T^6/\IZ_2 \times \IZ_2$. In this case the computation parallels those of the D6-D6 and D6-D8 systems, in the sense that the contributions coming from twisted sectors cancel each other and we end up with the intersection number $I_{ab}$ again given by (\ref{D8D8v}).

\subsection{Summary}

Let us summarize the above results on the chiral spectrum arising from D6-branes at angles and coisotropic D8-branes. Although now there are several possibilities in order to produce chirality from a pair of  D-branes, finding the chiral spectrum of a compactification always boils down to compute the intersection number $I_{ab}$ between two homology classes of 3-cycles $[\Pi_3^a]$ and $[\Pi_3^b]$. As advanced in (\ref{inter}) and (\ref{inter2}), such 3-cycles may correspond to those wrapped by a D6-brane $[\Pi_3^{D6}]$ or to the D6-brane charge which is dissolved into a coisotropic D8-brane $[\Pi_3^{D8}]$.

Because of the choice of discrete torsion taken on the $\T^6/\IZ_2 \times \IZ_2$ background, neither the D6-branes nor the D8-branes carry any twisted charge, and so the homological charges $[\Pi_3^{D6}]$ and $[\Pi_3^{D8}]$ can be understood in terms of (a subsector of) the $\T^6$ cohomology. This implies that the intersection numbers $I_{ab}$ can be expressed in terms of the wrapping and magnetic numbers (\ref{stackdata}) that define a D6-brane at angles or a coisotropic D8-brane in $\T^2 \times \T^2 \times \T^2$. In particular, they take the expressions (\ref{D6D6iii}), (\ref{D6D8iv}), (\ref{D8D8v}) or (\ref{D8D8orbiii}) depending on which particular pair of D-branes we are considering. 

Once these intersection numbers have been computed, everything works as in the case of D6-branes at angles discussed in \cite{csu01}. In particular, because we are performing the same orientifold projection that introduces O6-planes in our theory, we need to consider the D6's and D8's orientifold images. In the case of the D8-branes, this means that we need to add the images (\ref{finalprime}) in our compactification and that we must also compute the intersection numbers $I_{ab'}$. Finally, the D6$_a$-D6$_{a'}$ and D8$_a$-D8$_{a'}$ sectors may give rise to symmetric and/or antisymmetric chiral matter, and this part of the spectrum also depends on the intersection number $I_{a O}$ between $[\Pi_3^{D6}]$, $[\Pi_3^{D8}]$ and the O6-plane homology class, given by
\beqa \nonumber
[\Pi_3^{O6}] & = & [(1,0)\ (1,0)\ (1,0)] \ +\  [(1,0)\ (0,1)\ (0,-1)] \\
& + & [(0,1)\ (1,0)\ (0,-1)] \ +\  [(0,1)\ (0,-1)\ (1,0)].
\label{oriclass}
\eeqa
Such chiral spectrum is then summarized by table \ref{matter}.

\begin{table}[htb] \footnotesize
\renewcommand{\arraystretch}{1.25}
\begin{center}
\begin{tabular}{|c|c|}
\hline
\hspace{1cm} {\bf Sector} \hspace{1cm} &
\hspace{1cm} {\bf Representation} \hspace{1cm} \\
\hline\hline
$aa$   &  $U(N_a/2)$ vector multiplet  \\
 \hline
$ab+ba$   & $I_{ab}$ $(\fund_a,\antifund_b)$ chiral multiplets  \\
\hline
$ab^\prime+b^\prime a$ & $I_{ab'}$ $(\fund_a,\fund_b)$ chiral multiplets  \\
\hline
$aa^\prime+a^\prime a$ & $\frac 12 (I_{aa'} - 4 I_{aO}) \;\;
\Ysymm\;\;$ chiral  multiplets \\
          & $\frac 12 (I_{aa'} +  4 I_{aO}) \;\;
\Yasymm\;\;$ chiral multiplets \\
\hline
\end{tabular}
\end{center}
\caption{\small Gauge groups and chiral spectrum  spectrum for $\CN=1$ intersecting D6-branes and coisotropic D8-branes in the $\T^6/(\IZ_2\times \IZ_2)$ orientifold considered in this paper.}
\label{matter}
\end{table}

\section{Effective field theory\label{effective}} 

Having constructed BPS coisotropic D8-branes in $\T^6$ and in the $\T^6/\IZ_2 \times \IZ_2$ orientifold, we would now like to understand what is the effective field theory associated to them. In particular, we would like to derive the expressions for the F and D-terms that appear when the supersymmetry conditions (\ref{SUSYD8b}) are no longer met. As usual, these quantities can be extracted from the scalar potential generated in the $D=4$ effective theory and, more precisely, from the contribution of the D8-brane DBI action to such scalar potential. 

We have already encountered an effective field theory quantity in the previous section, namely the spectrum of $D=4$ chiral fermions arising from a D8-brane when intersecting others. In that case we found that everything depends on the D6-brane induced charge on the D8-brane, so one may wonder if this also true for any other quantity. We will see that this is indeed the case for some elements of the effective field theory, like the gauge kinetic functions, Fayet-Iliopoulos terms and massive $U(1)$'s of a D8-brane and that, in these cases, everything works like for a D6-brane wrapping a non-factorizable 3-cycle. On the other hand, there will be important differences when analyzing the $D=4$ superpotential. In particular, we will see that a open-closed superpotential will couple the D8-brane moduli to the K\"ahler moduli of the compactification, and that the generation of Yukawa couplings may or may not involve world-sheet instantons. 

Because our coisotropic D8-branes do not carry any RR twisted charge and we are working in the orbifold limit, all the effective theory quantities will depend on the toroidal D8-branes charges (\ref{final}) and on the closed untwisted moduli of the compactification. Recall that in the IIA orientifold that we are considering there are three untwisted K\"ahler moduli $T_i$, defined in terms of the complexified K\"ahler form $J_c$ as shown in (\ref{kahler}).\footnote{In this section we will express the real part of these K\"ahler moduli as $\re T_i = R_i^2 \tau_i$, where $R_i$ is the compactification radius in the $x^i$ direction (in units of $2\pi \sqrt{\a'}$) and $\tau_i$ has been defined in (\ref{holo2}).} The remaining moduli are the dilaton $S$ and the three complex structure moduli $U_i$, which can be encoded in the complexified 3-form \cite{gl04}
\beq
(4\pi^2 \a')^{3/2} \, \Omega_c  =   C_3 + i\re (C\Omega)
\label{omcdef}
\eeq
where in our conventions $C=e^{-\phi}$. The RR 3-form $C_3$ is to be expanded in a basis of 3-forms invariant under the orientifold action. In our case,
\beq
\Omega_c  =   iS \, dx^1 \wedge dx^2 \wedge dx^3 - i \sum_{i\not= j
  \not=k} U_i \, dx^i \wedge dy^j \wedge dy^k \  .
\label{ommod}
\eeq
It then follows that
\beq
\re S = e^{-\phi}  R_1 R_2 R_3  \quad ; \quad
\re U_i = e^{-\phi} R_i R_j R_k \tau_j \tau_k \quad , i\not= j \not=k  \ .
\label{sui}
\eeq
For future purposes we introduce the four-dimensional dilaton given by
\beq
e^{4\phi_4} = \frac{e^{4\phi}}{\CV^2} =  \left( \re S \, \re U_1 \, \re U_2 \, \re U_3\right)^{-1} 
\ ,
\label{dilaton4d}
\eeq
where $\CV=R_1^2 R_2^2 R_3^2 \tau_1 \tau_2 \tau_3$ is the volume of the internal $\T^6$ in string units.

The effective action for a $\D p$-brane is the sum of two terms. The first is the Dirac-Born-Infeld action given by
\beq
S_{\rm DBI} = -\mu_p \int_{\Sigma_{p+1}} d^{p+1} \xi \, e^{-\phi} \, 
\Tr \sqrt{\det(G+ B + 2\pi\a^\prime F)} \  ,
\label{dbi}
\eeq
where $\Sigma_{p+1}$ is the D-brane worldvolume. As usual, $G$ and $B$ are the pullbacks of the corresponding \deq10 tensors, whereas $F$ is the worldvolume field strength. The second term is the Chern-Simons action
\beq
S_{\rm CS} = -\mu_p \int_{\Sigma_{p+1}} \cc \wedge  
e^{\CF} \  ,
\label{csaction}
\eeq
where $\cc$ is a formal sum of RR forms. 

As usual, these quantities simplify for D-branes on tori or toroidal orbifolds. For instance, if we consider a D8$_a$-brane with data given by (\ref{final}), we can derive the useful result,
\beqa
\left. \det\left(G + \CF\right)\right|_{\Pi_5} \! \! \! \! & =  & \! \! \! \! (4\pi^2 \a^\prime)^5 
R_1^2 R_2^2 R_3^2 \left((n^i_a)^2 + (m^i_a \tau_i)^2\right) 
\big[(n_a^{yy}-n_a^{xx}\tau_j\tau_k)^2 + (n_a^{xy}\tau_j + n_a^{yx} \tau_k)^2 
\nonumber \\
& + & \frac{1}{R_j^2 R_k^2}\left|T_j T_k
- n_a^{xy} n_a^{yx} + n_a^{xx} n_a^{yy} \right|^2  \big]
\label{dbimod} 
\eeqa
where $|_{\Pi_5}$ stands for the pull-back on the 5-cycle that the D8-brane is wrapping. Observe that using the F and D-flatness conditions (\ref{genFterm}) and  (\ref{genDterm}) leads to
\beq
\int_{\Pi_5} \sqrt{\det\left(G + \CF\right)}\, =\, 
\int_{\Pi_5} \CF \wedge \re \Om  \, =\, 2\pi \a' \int_{\Pi_5} F \wedge \re \Om  
\label{susy4}
\eeq
where we have used that the $B$-field is a (1,1)-form and hence $B \wedge \Om = 0$. As we will shortly explain, moving away from these supersymmetry conditions allows instead to deduce F and D-terms in the scalar potential.

\subsection{Gauge coupling constants}

Let us consider a type II compactification, and a gauge theory arising from a $\D p$-brane wrapping a $(p-3)$-cycle $\Pi_{p-3}$ on the internal manifold. 
One basic question is which is the gauge coupling constant of such theory. As it is well known, this follows from reducing the Dirac-Born-Infeld action (\ref{dbi}) over the compact $\Pi_{p-3}$, and reading the gauge coupling from the resulting \deq4 action. In this way we obtain the general result \cite{johnson01}
\beq
\frac{2\pi}{g_p^2}\, =\, (2\pi)^{3-p} \a'^{\frac{3-p}{2}} \int_{\Pi_{p-3} }e^{-\phi} \sqrt{\det(G + \CF)}
\label{ggen}
\eeq

In an \neq1 supersymmetric theory it must be that $1/g_p^2$ is given by the real part of the holomorphic gauge kinetic function $f_p$. Furthermore, the imaginary part of $f_p$ determines the axionic couplings. These couplings can be derived from reducing the Chern-Simons action (\ref{csaction}) over $\Pi_{p-3}$ and so, eventually, $\im f_p$ can be also deduced. For instance, for a D6-brane, $\im f_6 \sim \int_{\Pi_3} C_3$, whereas for a coisotropic D8-brane,
\beq
\im f_8  = -\frac{1}{2\pi} (4\pi^2 \a')^{-3/2} \int_{\Pi_5} C_3 \wedge \frac{F}{2\pi} \  .
\label{imf8}
\eeq
Clearly, $\im f_8$ depends linearly on the imaginary part of the
complex structure moduli that come from the RR 3-form. On the other
hand, $\re f_8$ seems to have a complicated dependence on the NSNS
piece of the moduli. However, once the supersymmetric
conditions are used we can easily reconstruct the holomorphic gauge
function.

Indeed, substituting eq.(\ref{susy4}) in the gauge coupling constant gives
\beq
\re f_8\, =\, \frac{1}{g_8^2}=\frac{1}{2\pi} (4\pi^2 \a')^{-3/2} \int_{\Pi_5} \frac{F}{2\pi} \wedge \re(e^{-\phi} \Om) \  .
\label{ref8}
\eeq
so combining with the imaginary part yields
\beq
f_8 \, =\, -\frac{i}{2\pi} \int_{\Pi_5} \frac{F}{2\pi} \wedge \Om_c \, =\, 
 -\frac{i}{2\pi} \int_{[\Pi_3]} \Om_c
\label{fullf8}
\eeq
where $[\Pi_3]$ is the Poincar\'e dual of $[F/2\pi]$ in $\Pi_5$. It is thus clear that the gauge kinetic function only depends on the D6-brane charge induced on the coisotropic D8-brane. 

In particular, in our toroidal compactification we can use (\ref{ommod}) to express $\Omega_c$ in terms of the moduli, so that for the $\D 8_a$-brane (\ref{final}) we obtain
\beq
f_{8a} = - n_a^i( n_a^{yy} S - n_a^{xx} U_i) - m_a^i (n_a^{yx} U_j + n_a^{xy} U_k)
\label{fmoduli}
\eeq
which explicitly shows that $f_{8a}$ is a holomorphic function of the dilaton and the complex structure moduli. In addition, using the dictionary (\ref{iden}) one can check that this expression matches the one obtained in \cite{cim02a} for intersecting D6-branes.

\subsection{Scalar potential}

Basically, the contribution of a $\D p$-brane to the scalar potential of an effective theory can be computed from its tension, as derived from the Dirac-Born-Infeld action. For example, for a coisotropic D8-brane we have
\beq
V_{\D 8}= \mu_8 \int_{\Pi_5} e^{4\phi_4} e^{-\phi} \sqrt{\det(G+\CF)} \ . 
\label{vd8g}
\eeq
where the factor $e^{4 \phi_4}$ appears because the action term is actually $\int d^4x \sqrt{\tilde g} \, V_{\D 8}$, $\tilde g_{\mu \nu}$ being the 4-dimensional metric in Einstein frame. In the toroidal compactification the four-dimensional dilaton is given by (\ref{dilaton4d}).

To proceed further, we consider a $\D 8_a$ brane for which the Dirac-Born-Infeld
determinant is computed in (\ref{dbimod}). As in \cite{martucci06}, it is convenient to define D-term and F-term densities as
\beqa
e^{-\phi} \re (\CF \wedge \Omega) \left|_{\Pi_5}\right. & = & 
(4 \pi^2 \a^\prime)^{5/2} \,  \Theta_a \, d{\rm vol}_{\Pi_5} \ , 
\nonumber \\ 
e^{-\phi} \im (\CF \wedge \Omega) \left|_{\Pi_5}\right. & = & 
(4 \pi^2 \a^\prime)^{5/2} \, \CD_a \, d{\rm vol}_{\Pi_5} \ , 
\label{dfdensities} \\ 
(\CF + iJ)^2 \left|_{\T^4}\right. & = & (4 \pi^2 \a^\prime)^2 \, \CQ_a \, d{\rm vol}_{\T^4} \ . 
\nonumber 
\eeqa
These quantities have been basically computed in (\ref{genFterm}) and (\ref{genstable}). Explicitly,
\beqa
\Theta_a & = & \left[
 - n_a^i( n_a^{yy} \re S - n_a^{xx} \re U_i) - m_a^i (n_a^{yx} \re U_j + n_a^{xy} \re U_k)  \right]   \ , 
\nonumber \\[0.2cm] 
e^{2\phi_4} \CD_a  & = & \left[ 
- \frac{ m_a^i n_a^{yy}}{\re U_i} + \frac{ m_a^i n_a^{xx}}{\re S} 
+ \frac{ n_a^i n_a^{xy}}{\re U_j} + \frac{ n_a^i n_a^{yx}}{\re U_k}  \right] \equiv D_a  \ , 
\label{fulldensities} \\[0.2cm]  
\CQ_a & = &  \left[T_j T_k - n_a^{xy} n_a^{yx} + n_a^{xx} n_a^{yy} \right] \ , 
\nonumber 
\eeqa
where we have also used (\ref{sui}) and (\ref{dilaton4d}).
Notice that $\Theta_a = \re f_{8\,a}$.  

The next step is to identify the terms in (\ref{dbimod}) to arrive at
\beq
V_{\D 8_a}= \mu_8 (4 \pi^2 \a^\prime)^{5/2} \, 
e^{4\phi_4} \sqrt{\Theta_a^2 + \CD_a^2 + e^{-2\phi} ||\ell_i||^2 |\CQ_a|^2} \ , 
\label{vd8}
\eeq
where $||\ell_i||^2= 4\pi^2 \alpha^\prime R_i^2  \left((n^i_a)^2 + (m^i_a \tau_i)^2\right)$.
Expanding the square root readily gives
\beq
V_{\D 8_a}= \frac{1}{8\pi^3 \a^{\prime 2}} \, 
e^{4\phi_4} \left\{|\Theta_a| + \frac{\CD_a^2}{2 |\Theta_a|} + 
\frac{e^{-2\phi} ||\ell_i||^2 |\CQ_a|^2}{2 |\Theta_a|} + \cdots \right\}
\label{vd8exp}
\eeq
This expansion supports the interpretation of the supersymmetry constraints
$\im (\CF \wedge \Omega)=0$ and $(\CF + iJ)^2 =0$ as D-flatness and F-flatness conditions. On the other hand, imposing the supersymmetry conditions leaves only the first term in (\ref{vd8exp}). In fact, in a supersymmetric configuration this term will cancel against the contribution of the orientifold planes given by
\beq
V_{ori} = -16  e^{4\phi_4} e^{-\phi} ||\ell_{ori}|| \ .
\label{Vori}
\eeq
In the $\T^6/\IZ_2 \times \IZ_2$ orientifold that we are considering
\beq
||\ell_{ori}|| = R_1 R_2 R_3 (1 + \tau_2 \tau_3  + \tau_3 \tau_1  + \tau_1 \tau_2)  \ .
\label{lori}
\eeq
and so it is easy to check that the RR tadpole cancellation conditions (\ref{tadpoled6}) imply the vanishing of the full potential of a supersymmetric brane setup.
This is the usual statement that, by supersymmetry, NSNS tadpoles will also 
cancel in this compactification.

In the scalar potential (\ref{vd8exp}) we clearly identify a D-term appearing in the standard supergravity form $V_D=\frac12 D_a D^a$, where $D_a = e^{2\phi_4} \CD_a$ up to normalization, and the index $a$ is raised with the gauge metric $(\re f_{8\, a})^{-1}$. In the case of D6-branes, the analogous D-term corresponds to Fayet-Iliopoulos terms associated to the D-branes $U(1)$ factors \cite{km99,cim02a} (see also \cite{bbkl02,vz2}). It is clear that the similar result should hold here, since we could obtain (\ref{fulldensities}) by replacing the magnetized D8-brane by a non-factorizable D6-brane whose wrapping numbers are given by (\ref{iden}).

On the other hand, the scalar potential also includes an F-term contribution, namely
\beq
V_F = \frac{1}{16\pi^3 \a^{\prime 2}} \, 
\frac{e^{2\phi} ||\ell_i||^2 |\CQ_a|^2}{\CV^2 \re f_{8a}}  \  .
\label{vd8F}
\eeq
In the following we will argue that this F-term is due to a superpotential
\beq
W = \frac{\sqrt2}{\pi^2 (\a^\prime)^{3/2}} \,  X \CQ_a  \ ,
\label{luis}
\eeq
where $\CQ_a$ depends on closed K\"ahler moduli as shown in (\ref{fulldensities}),
whereas $X$ is an open string modulus of the coisotropic $\D 8_a$-brane.
The task is to show that $V_F$ is of the standard form 
\beq
e^{K} K^{\bar X X} \left| \frac{\partial W}{\partial X} \right|^2  \ ,
\label{VFsugra}
\eeq
where $K$ is the K\"ahler potential of the closed and open moduli and,
as usual, $K^{\bar A B}$ is the inverse of $K_{\bar A B} =\partial_{\bar A} \partial_B K$.

First, the K\"ahler potential has the general structure
\beq
K = \hat K + \tilde K_{X \bar X} X \bar X + \cdots  \ ,
\label{kpotgen}
\eeq
with $\hat K$ and $\tilde K_{X \bar X}$ depending only on the closed moduli. For  $\hat K$ we have the well-known tree-level result
\beq
\hat K = -\log(S+\bar S) - \sum_{i=1}^3 \log(U_i+\bar U_i) - 
\sum_{i=1}^3  \log(T_i+\bar T_i)
\label{ktree}
\eeq
and so, to first order in $X$,
\beq
e^K = \frac{e^{4\phi}}{128\CV^3} + \cdots  \ .
\label{expK}
\eeq
Second, we need to determine $\tilde K_{X \bar X}$, and our strategy will be to compute the kinetic energy of the open string modulus, i.e. $\tilde K_{X \bar X} \partial_\mu X \partial^\mu \bar X$, from the Dirac-Born-Infeld action.

Let us first identify the open string modulus $X$. To simplify the discussion we will consider a fractional 
$\D 8_a$-brane wrapping the 5-cycle $(1,0)_1 \times ((\T^2)_2  \times (\T^2)_3)/\IZ_2$ (see section 3.2). 
Then, the obvious moduli of our D8-brane in the $\D 8_a \D 8_a$ sector are the position $y^1$ of $(1,0)_1$ 
in $(\T^2)_1$, and the Wilson line $A_1$ along the unique 1-cycle $(1,0)_1$. 
These two real moduli can be arranged into a complex field as
\beq
X = R_1^2 \tau_1 \, y^1 + i A_1
\label{xmod}
\eeq
where the factor of $\re T_1$ in $\re X$ is necessary in order to obtain
a canonical kinetic energy. In fact, upon reducing the Dirac-Born-Infeld action
the terms quadratic in space-time derivatives of $y^1$ and $A_1$, in Einstein frame, turn out to be
\beq
\frac{1}{4\pi \a^\prime} \, \frac{e^{2\phi}}{\CV} \, \sqrt{\tilde g} \, \re f_{8a} \left(R_1^2 \tau_1^2 \partial_\mu y^1  \partial^\mu y^1 +  
R_1^{-2} \partial_\mu A_1  \partial^\mu A_1 \right)
\label{kinterms}
\eeq
where the factor of $\re  f_{8a}$ basically comes from the integral over the 5-cycle that can be performed using (\ref{dbimod}) plus the supersymmetry constraints.

Then, from the kinetic terms we obtain
\beq
\tilde K_{X \bar X} = \frac{1}{4\pi \a^\prime} \,
\frac{e^{2\phi}}{\CV ||\ell_i||^2} \, \re f_{8a}  \  ,
\label{tildeK}
\eeq
since, in the case at hand, $||\ell_i||^2 = R_1^2$. 
This is enough to check that the scalar potential $V_F$  has the expected
form (\ref{VFsugra}). Indeed, collecting previous results we find
\beq
e^{K} K^{\bar X X} =  \frac{\pi \a^\prime}{32} \, \frac{e^{2\phi} ||\ell_i||^2}{\CV^2 \re f_{8a}} + \cdots  \ ,
\label{VFexp}
\eeq
to first order in $X$. Finally, the proposed superpotential obviously
verifies $\frac{\partial W}{\partial X} \sim \CQ_a$.

To summarize, we find that for a coisotropic D8-brane (\ref{final}) the worldvolume flux $F$ induces a superpotential linear in its open string modulus $X_i$ which, as in the case of D6-branes, is a combination of transverse position and Wilson line in $(\T^2)_i$. Such superpotential also involves the K\"ahler moduli of the compactification, and has the general form
\beq
W\, =\, X_i (T_jT_k - n)
\label{super}
\eeq
where $n$ comes from integrating $F^2$ over $(\T^2)_j \times (\T^2)_k$. Such type of superpotentials have also been obtained in the case of magnetized D7-branes \cite{lmrs,jl}.

In general, the existence of a open-closed string superpotential is quite an interesting feature of a D-brane model and, in principle, it may provide new sources of moduli stabilization. Naively, one would say that in the presence of a superpotential of the form (\ref{super}) the K\"ahler moduli should be fixed as $T_jT_k = n$, which is precisely the F-flatness condition for our D8-brane. One should be however careful, because there could be extra superpotential terms involving $X_i$ and then this naive picture could be complicated. For instance, we have analyzed the open string modulus $X_i$, which exists for any fractional D8$_a$-brane on $(n_a^i,m_a^i)_i \times ((\T^2)_j  \times (\T^2)_k)/\IZ_2$. However, there could be extra open string moduli, say $X_j$ and $X_k$, coupling as
\beq
W_X\, =\, X_i X_j X_k
\label{superadj}
\eeq
as is the case for D6-branes \cite{douglas98,csu01}. Then the F-flatness condition for $X_i$ would read
\beq
\frac{\p (W + W_X)}{\p X_i} \, = \, (T_jT_k + X_j X_k - n) \, =\, 0
\label{Xiflat}
\eeq
and hence the deviation of the K\"ahler moduli from the BPS condition could be compensated by giving a vev to open string moduli. Notice that this phenomenon is quite analogous to the case of D-terms for D6-branes, where the scalar potential does not fix the complex structure moduli of the compactification, but only a combination of closed and open string moduli. In this sense, the fact that $X_jX_k$ take a vev could be interpreted as some kind of D8-brane recombination.

Notice also that, as a byproduct of our analysis, we have determined the K\"ahler
metric of the open string field $X$. This metric has a very simple expression
as function of the closed moduli. For instance, for a 
$\D 8_a$-brane wrapping $(1,0)_1 \times (\T^2)_2  \times (\T^2)_3$
we find
\beq
\tilde K_{X \bar X} = 
\frac{1}{4\pi \a^\prime \re T_1}\left(\frac{n_a^{xx}}{\re S} - \frac{n_a^{yy}}{\re U_1}  
\right)  \ .
\label{tildeK1}
\eeq
If the 5-cycle is instead $(0,1)_1 \times (\T^2)_2  \times (\T^2)_3$
the metric is given by 
\beq
\tilde K_{X \bar X} = - 
\frac{1}{4\pi \a^\prime\re T_1}\left(\frac{n_a^{xy}}{\re U_2} + \frac{n_a^{yx}}{\re U_3}  
\right)  \ .
\label{tildeK2}
\eeq 
In general, we just need to apply eq.(\ref{tildeK}). For magnetized D7-branes
the open moduli metric has been discussed in \cite{lmrs1,lss,fi,lmrs,jl}.

\subsection{Massive $U(1)$'s}

The axionic coupling (\ref{imf8}) not only enters in the gauge kinetic function of the effective theory, but it is also an important ingredient in the cancellation of $U(1)$ anomalies through a generalized Green-Schwarz mechanism \cite{iru98}. One can see that the imaginary parts of the dilaton ($S\equiv U_0$) and the complex structure moduli couple to the $D=4$ gauge field strengths as
\beq
\sum_{L=0}^3 d_L^a \im U_L \Tr (F_a \wedge F_a) \ .
\label{dcoef}
\eeq
For instance, if $F_a$ arises from the coisotropic $\D8_a$, with data given in (\ref{final}) and $i=1$, we find the coefficients 
\beq
d_L^a=(-n_a^1 n_a^{yy},\    
-n_a^1 n_a^{xx}, \   
m_a^1 n_a^{yx}, \   
m_a^1 n_a^{xy})  \ .
\label{dcoefs}
\eeq

In order to cancel $U(1)$ anomalies we also need  $B \wedge \Tr F_a$ couplings, which can also be obtained from the Chern-Simons action. Indeed, from (\ref{csaction}) we obtain couplings of the form 
\beq
\Tr F_a \wedge \int_{\Pi_5^a} \CF \wedge C_5 = \sum_{L=0}^3 c_L^a B_L \wedge \Tr F_a \ .
\label{massterm}
\eeq
where the $B_L$ are 2-forms dual to the 0-forms $\im U_L$.
In the example above ($i=1$) the coefficients $c_L^a$ are given by
\beq
c_L^a=(-N_a m_a^1 n_a^{xx},\    
-N_a m_a^1 n_a^{yy}, \   
N_a n_a^1 n_a^{xy}, \   
N_a n_a^1 n_a^{yx})  \ .
\label{ccoefs}
\eeq
Obviously, these couplings exist only for Abelian $U(1)$ factors. 

The couplings (\ref{dcoef}) and  (\ref{massterm}) give a contribution
\beq
\sum_L c_L^i d_L ^a \  , 
\label{u1anomaly}
\eeq
to mixed and cubic $U(1)_a$ anomalies. Once the RR tadpole conditions (\ref{tadpoled6}) have been imposed, this piece 
will cancel the remaining triangle anomalies due to the charged massless fermions in the spectrum. 
In fact, the $U(1)$-$SU(N)^2$ anomaly cancellation works exactly like in the case of D6-branes, so we refer the 
reader to the original references \cite{afiru00,imr01} for a detailed computation.

As usual, when the terms $B \wedge \Tr F_a$ do not vanish, there will be combinations of $U(1)$ factors that acquire a mass. A generic $U(1)$ generator
$Q=\sum_i \xi_i Q_i$, where $Q_i$ arises from the $i^{th}$ stack, will remain
massless provided that
\beq
\sum_i c_L^i \xi_i = 0  \quad ; \quad \forall L
\label{masscond}
\eeq   
which guarantees that the corresponding $U(1)$ does not couple to any RR 2-forms $B_L$.

\subsection{Yukawa couplings}

\begin{figure}
\epsfysize=3.5cm
\begin{center}
\leavevmode
\epsffile{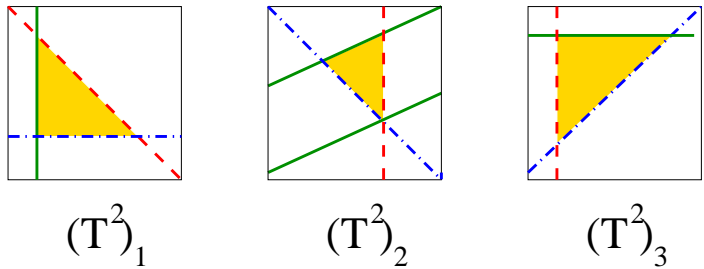}
\end{center}
\caption{Yukawa couplings in intersecting D6-branes involve
a sum over triangle shaped instantons.}
\label{d6d6d6}
\end{figure}

Other effective field theory quantities of obvious interest are the Yukawa couplings among chiral fields. In the case of intersecting D6-branes, the flavor-dependent contribution to the Yukawa couplings is given by a sum over worldsheet instantons \cite{afiru00} (see fig. \ref{d6d6d6}). The couplings among three chiral fields labeled $\alpha, \beta, \gamma$, have the qualitative form
\beq
Y_{\alpha\beta\gamma} \ \propto \ 
\prod_{i} \left( \sum_{k_i \in \IZ} e^{-\frac {A_{\alpha\beta\gamma}(k_i)}{2\pi \alpha '}} \right)
\label{yuk6}
\eeq
where $i=1,2,3$, labels the three tori and $A_{\alpha\beta\gamma}(k_i)$ 
is the area of the triangular instantons on each of them.
This sum was shown in \cite{cimyuk} to be equal to a product of
Jacobi theta-functions in the toroidal/orbifold case. 
The type I mirror of these orientifolds consists of D5-branes and magnetized D9-branes, and the Yukawa couplings come from  
a pure field-theoretical calculation.  They are given by
the overlap integral over the 6 extra dimensions of the three wave functions 
of the zero modes $\alpha, \beta, \gamma$, i.e.,
\beq 
Y_{\alpha\beta\gamma}\ \propto \
\int\raisebox{-2.5ex}{}_{\!\!\!\!\!\! (\T^2)_1\times (\T^2)_2\times (\T^2)_3}
\!\!\!\!\!\!\!\!\!\!\!\!\!\!\!\!\!\!\!\!\!\!\!\!\!\!\!\!\!\!\!\!\!\!\!\!
d^6y \ \Psi_\alpha(y)\Psi_\beta(y)\Psi_\gamma(y)
 \label{yukmag}
\eeq
where some of the wavefunctions are delta functions when D5-branes are involved in the Yukawa coupling. As checked in \cite{cimmag}, this computation yields the expected T-dual of the computation in terms of intersecting D6-branes.

In the presence of coisotropic D8-branes we find three 
new different classes of brane configurations giving rise to Yukawa couplings among chiral fields. In these cases, the origin of Yukawa couplings turns out to be a combination of the above two mechanisms. 
The three new classes are as follows:

\begin{itemize}

\item [{\it a)}] {\it Yukawa couplings involving two $\D 6$'s and one $\D 8$.}

\begin{figure}
\epsfysize=5cm
\begin{center}
\leavevmode
\epsffile{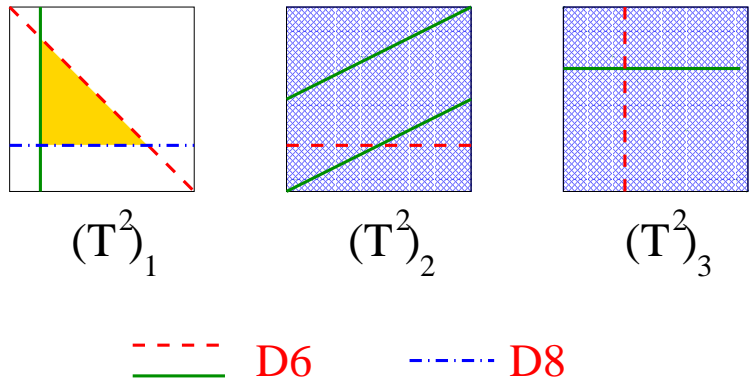}
\end{center}
\caption{Yukawa couplings involving two D6-branes and one D8 
contain contributions both from instantons (first torus in the figure) 
and integrals over overlapping wave functions.}
\label{d6d6d8}
\end{figure}

This is depicted in fig.~\ref{d6d6d8} for the case of a D8-brane with
the 1-cycle in the first torus. The Yukawa coupling will have 
two factors, one  contribution coming from a sum over triangle instantons in the
first torus (see the figure) and the other from the overlap integral 
of wave functions over the second and third torus. One thus has a structure for
the couplings of the form
\beq
Y_{\alpha\beta\gamma} \ \propto \
\left( \sum_{k_1} e^{-\frac {A_{\alpha\beta\gamma}(k_1)}{2\pi \alpha '}} \right)
\times
\int\raisebox{-2.5ex}{}_{\!\!\!\!\!\! (\T^2)_2\times (\T^2)_3}\!\!\!\!\!\!\!\!\!\!\!\!\!\!\!\!\!\!\!\!\!\!
d^4y \ \Psi_\a(y) \Psi_\beta(y)\Psi_\gamma(y)
\label{yuk668}
\eeq
where the wavefunctions have only support in the D-brane intersections. In particular, the wavefunction arising from the D6D6 sector will be a delta function on $(\T^2)_2\times (\T^2)_3$.

\item  [{\it b)}]  {\it Yukawa couplings involving two $\D 8$'s and one $\D 6$}.

\begin{figure}
\epsfysize=7cm
\begin{center}
\leavevmode
\epsffile{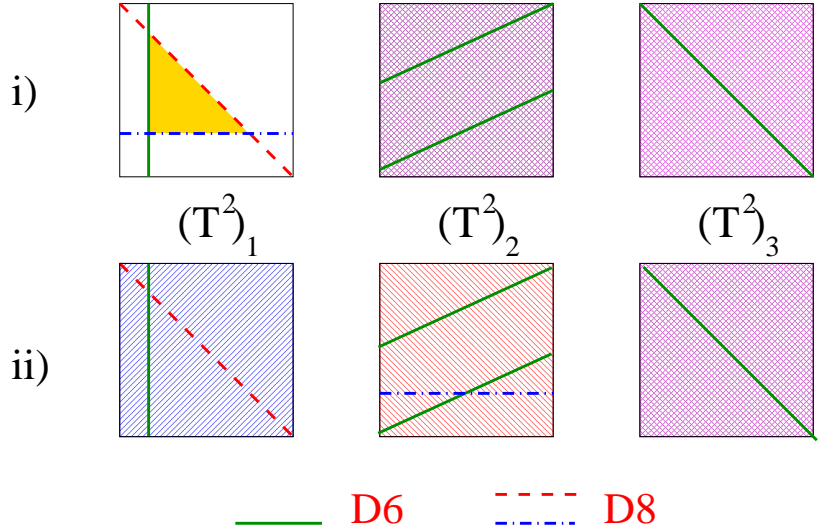}
\end{center}
\caption{ The two general classes of Yukawa couplings 
in configurations with two D8's and one D6. In the first one
there are contributions from a sum over instantons and
overlap integrals. In the second case there is no sum over 
instantons.}
\label{d6d8d8}
\end{figure}

This is depicted in fig.~\ref{d6d8d8}. As shown in the figure, there are
two classes of  configurations of this type. In the first the 1-cycle 
of both D8-branes are in the same torus, i.e. $(\T^2)_1$. In this case the 
structure of the Yukawa couplings is similar to the previous case and the
Yukawa coupling has the general form in eq.(\ref{yuk668}). In the second case in the
figure there is no sum over instantons and the structure is rather as in 
eq.(\ref{yukmag}). 

\item  [{\it c)}] {\it Yukawa couplings involving three $\D 8$'s}.

\begin{figure}
\epsfysize=9cm
\begin{center}
\leavevmode
\epsffile{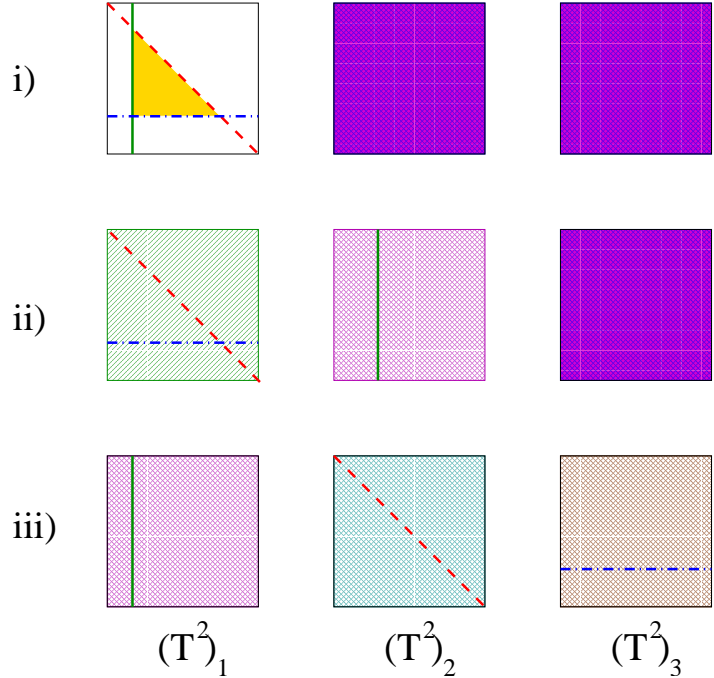}
\end{center}
\caption{The three general classes of Yukawa couplings 
involving  three D8's.}
\label{d8d8d8}
\end{figure}

In this case there are three possible configurations of the D8's, as
shown in  fig.~\ref{d8d8d8}. In the first configuration again we
have a sum over instantons in one of the tori ($(\T^2)_1$ in the figure)
and the Yukawa coupling will be given by an expression as in 
eq.(\ref{yuk668}). In the other two cases we will have an expression as in
eq.(\ref{yukmag}).

\end{itemize}

It would be interesting to see which kind of textures yield these new classes of Yukawa couplings. However, a detailed computation of these Yukawa couplings goes beyond the scope of this paper, and is left for future work.

\section{Coisotropic model building \label{modelitos}}

The coisotropic D8-branes discussed in the previous sections
turn out to be interesting from the model building point of view.
One of the reasons is that coisotropic D8-branes have induced
D6-brane charges corresponding to non-factorizable 3-cycles.
This allows for new ways to cancel RR tadpoles in constructions
giving rise to a MSSM-like spectrum.  In addition the presence of
an appropriate set of coisotropic D8-branes
give rise to superpotential couplings involving the K\"ahler
moduli and may be useful in order to fix them.
We postpone for future work
a systematic analysis of the model-building possibilities of
this new tool, but we will present here two examples
of  $\IZ_2 \times \IZ_2$ orientifold 
models  with both D8- and D6-branes and a chiral spectrum
close to that of a 3-generation MSSM-like model.
In the first example the MSSM fields will reside on a set
of intersecting D6-branes and the addition of a stack of
coisotropic D8-branes will ensure RR tadpole cancellations. 
In the second example the MSSM fields will reside at the intersection 
of both D8- and D6-branes and extra coisotropic D8-branes 
will be added both to cancel tadpoles and provide superpotential
couplings for the untwisted K\"ahler moduli.

\subsection{An MSSM-like model}

The first model will be based on one of the `triangle quiver' 
toroidal models discussed in section 4.4 of ref.\cite{cim02b},
concretely, that with $\beta^2=1$ (see that reference for details).
In this model the third torus is not a square but a `tilted'
torus. The extension of this set of D6-branes to the
$\IZ_2 \times \IZ_2$ orientifold case is simply done by doubling the 
number of D6-branes as in \cite{csu01}. We thus have D6-branes 
with wrapping numbers as in table \ref{mssm1}. 
\begin{table}[htb] \footnotesize
\renewcommand{\arraystretch}{1.25}
\begin{center}
\begin{tabular}{|c||c|}
\hline $N_\a$\  D$p_\a$ &
D8$_\a:\ (n^i_\a,m^i_\a)_i\times (n^{xx}_\a,n^{xy}_\a,n^{yx}_\a,n^{yy}_\a)_{jk} $ \\
 & D6$_\a:\  (n^1_\a,m^1_\a)_1 (n^2_\a,m^2_\a)_2 (n^3_\a,m^3_\a)_3 $ \\
 \hline\hline
$N_a=6$\ D6$_a$ & $(1,0)_1  (3,1)_2  (3,-1/2)_3$ \\
\hline
$N_b=4$\ D6$_b$ & $(1,1)_1(1,0)_2(1,-1/2)_3 $ \\
\hline
  $N_c=2$\ D6$_c$  & $(0,1)_1(0,-1)_2(2,0)_3$ \\
\hline
$N_d=2$\ D6$_d$ & $(1,0)_1 (3,1)_2 (3,-1/2)_3$ \\
\hline\hline
 $N_M=4$ \ D8$_X$ & $  (-1,-3,-2,-5)_{12}\times (-3,1/2)_3 $ \\
 \hline \end{tabular}
\end{center}
\caption{\small  Set of D6 and  coisotropic D8-branes giving rise to
the MSSM-like model in the text.}
\label{mssm1}
\end{table}

One can easily check that the set of D6-branes in this model 
are not enough to cancel RR-tadpoles. In particular, their
contribution to each of the 4 tadpoles is respectively
$(76,-4,-2,-4)$ whereas that of the orientifold planes 
is $(-16,8,8,16)$ \footnote{Note that the second and third 
orientifold charges are halved due to the tilting of the 
third torus.}. In order to cancel tadpoles we need 
branes contributing charges as $(-60,-4,-6,-12)$. 
A very economical possibility is to add a stack of
4 coisotropic D8-branes $M$ (and their corresponding 
images under the orientifold operation)
as shown in the table. One can easily check that 
indeed this simple addition cancels all tadpoles.
All D-branes in the model preserve the same \neq1 SUSY for appropriate
choices of the complex structure moduli. 
In particular the D-term conditions are 
\beq
2\tau_1=2\tau_2=\tau_3 \ \ \ ; \ \ 
\tau_1\tau_2\tau_3=18\tau_1+12\tau_2+5\tau_3
\eeq
which are obeyed for $\tau_1=\tau_2=\tau_3/2=\sqrt{20}$. 
A superpotential of the 
form 
\beq
W_{D8} \ =\ X_3(1-T_1T_2)
\eeq
is created due to the presence of the D8-brane. In absence of the superpotential (\ref{superadj}) or for fixed extra D8-brane moduli, this would constrain the $T_{1,2}$ K\"ahler moduli to satisfy $T_1T_2=1$.
\begin{table}[htb] \footnotesize
\renewcommand{\arraystretch}{1.3}
\begin{center}
\begin{tabular}{|c|c|c|c|c|c|c|c|c|}
\hline Intersection &
 Matter fields  &   &  $Q_a$  & $Q_b $ & $Q_c $ & $Q_d$ & $Q_M$  & $Q_Y$ \\
\hline
\hline $ab$ & $Q_L$ &  $(3, 2)$ & 1  & -1 & 0 & 0 &0& 1/6 \\
\hline $ab'$  & $q_L$ & $2(3,2)$ &  1  & 1  & 0  & 0 & 0   & 1/6 \\
\hline $ac$ & $U_R$ &  $3( {\bar 3},1)$ &  -1  & 0  & 1  & 0 & 0  & -2/3 \\
\hline $ac'$  & $D_R$   &  $3({\bar 3},1)$ &  -1  & 0  & -1  & 0 & 0  & 1/3 \\
\hline $bd$ & $ L $ &  $(1,2)$ &  0   & -1 & 0  & 1 & 0&  -1/2 \\
\hline $bd'$ & $ l $ &  $2(1,2)$ &  0  & 1   & 0  & 1 & 0 & -1/2 \\
\hline $cd$ & $N_R$ & $3(1,1)$ &  0  & 0  & 1  & -1  & 0 &  0   \\
\hline $cd'$ & $E_R$  & $3(1,1)$ &  0  & 0  & -1  & -1  & 0 & 1 \\
\hline $bc$ &  $H$ & $(1,2)$ &  0 & -1 & 1  &  0 & 0 & -1/2 \\
\hline $bc'$ & $ {\bar H}$ &  $(1,2)$ & 0 & -1 & -1 & 0 & 0
& 1/2 \\
\hline\hline
$bM$ & $ {F}$ &  $ 3(1,2;2_M)$ & 0 & -1 & 0 & 0 & 1 & 0 \\
\hline
$bM^*$ & $ {\bar F}$ &  $ 2(1,2;2_M)$ & 0 & 1 & 0 & 0 & 1 & 0 \\
\hline
$cM$ & $ {G}$ &  $ 5(1,1;2_M)$ & 0 & 0 & -1 & 0 & 1 & 1/2 \\
\hline
$cM^*$ & $ {\bar G}$ &  $ 5(1,1;2_M)$ & 0 & 0 & 1 & 0 & 1 & -1/2 \\
\hline
\end{tabular}
\end{center}
\caption{\small Chiral spectrum of the MSSM-like model
discussed in the text. The hypercharge generator is defined as
$Q_Y = \frac 16 Q_a - \frac 12 Q_c - \frac 12 Q_d$. \label{espectrosm}}
\end{table}

The gauge group is initially $U(3)_c\times U(2)_L\times U(1)_{B-L}\times U(1)_R
\times U(2)_M$. However, three out of the 5 $U(1)$'s are anomalous
and only the hypercharge and an additional $U(1)$ (the one characteristic of
left-right symmetric models) survives the $B \wedge F$ couplings described in the last section, remaining as massless $U(1)$'s. The chiral spectrum is shown in table \ref{espectrosm} and consists of the content of the MSSM plus some additional SM doublets and
singlets. Further antisymmetric and symmetric $U(2)_M$ chiral fields
uncharged under the SM exist which we do not display. 

This model, which is of phenomenological interest by itself, 
exemplifies how coisotropic D8-branes may be a useful tool
for model-building purposes. In this example the role of
the D8 was auxiliary, in the sense that it helped in
cancelling RR tadpoles and restricting K\"ahler moduli
but the MSSM fields reside on D6-branes. Models
analogous to this in which MSSM fields live on D8-branes
can also be constructed. This is illustrated in the
next example in which also additional D8-branes are appended in
order to provide superpotential couplings for the 
untwisted K\"ahler moduli.

\subsection{A left-right symmetric MSSM-like model}

In our second example
the brane structure is reminiscent of the MSSM-like D6-brane configuration
introduced in \cite{cimDESY,cimyuk} and first included in a 
tadpole free global $\IZ_2\times \IZ_2$ orientifold compactification
in \cite{ms04}. Consider a D8-D6 brane configuration as listed in
table \ref{mssmlr}.
\begin{table}[htb] \footnotesize
\renewcommand{\arraystretch}{1.25}
\begin{center}
\begin{tabular}{|c||c|}
\hline $N_\a$\  D$p_\a$ &     
D8$_\a:\ (n^i_\a,m^i_\a)_i\times (n^{xx}_\a,n^{xy}_\a,n^{yx}_\a,n^{yy}_\a)_{jk} $ \\
 & D6$_\a:\  (n^1_\a,m^1_\a)_1 (n^2_\a,m^2_\a)_2 (n^3_\a,m^3_\a)_3 $ \\
 \hline\hline 
$N_a=6+2$\ D8$_a$ & $(1,0)_1\times (1,3,-3,-10)_{23}$ \\
\hline
$N_b=2$\ D6$_b$ & $(0,1)_1(1,0)_2(0,-1)_3 $ \\
\hline
  $N_c=2$\ D6$_c$  & $(0,1)_1(0,-1)_2(1,0)_3$
\\ \hline\hline
  $N_M=4$ \ D6$_M$  & $(-2,1)_1(-3,1)_2(-3,1)_3$\\
\hline
 $N_X=2$ \ D8$_X$ & $ (1,0)_2 \times (1,0,0,-2)_{31} $ \\
\hline
 $N_Y=2$\  D8$_Y$ & $  (1,0)_3 \times (1,0,0,-2)_{12} $ \\
 \hline \end{tabular}
\end{center}
\caption{\small  Set of D6 and  coisotropic D8-branes giving rise to 
the left-right symmetric MSSM-like model in the text.}
\label{mssmlr}
\end{table}

The stacks of branes $a$, $b$, $c$,  contain the SM gauge group and particles 
whereas the stacks  $M$, $X$, and  $Y$  are auxiliary branes whose mission is
helping in cancelling tadpoles. In addition the D8's   $X$ and $Y$ also
contribute to the fixing of the three untwisted K\"ahler moduli $T_i$.
Note that the coisotropic D8-branes involved in the model have
induced D6-charge given by a sum of two factorized cycles:
\beqa
\D 8_a&:& \ ( {1,0})_1\times ( {1,3,-3,-10})_{23}
= ( {1,0})_1\times [( {3,1})( {3,-1})+( {1,0})( {1,0})]\nonumber\\
\D 8_X&:& \ ( {1,0})_2\times ( {1,0,0,-2})_{31}
= ( {1,0})_2\times [( {0,1})( {0,-1})+( {2,0})( {1,0})]\nonumber\\
\D 8_Y&:& \ ( {1,0})_3\times ( {1,0,0,-2})_{12}
= ( {1,0})_3\times [( {0,1})( {0,-1})+( {2,0})( {1,0})]\nonumber
\eeqa
\begin{table}[htb] \footnotesize
\renewcommand{\arraystretch}{1.3}
\begin{center}
\begin{tabular}{|c|c|c|c|c|c|}
\hline Intersection &
 Matter fields  &   &  $Q_a$ & $Q_d$ & $Q_M$   \\
\hline
\hline $ab$ & $Q_L$ &  $3(3,2,1)$ & 1  & 0 & 0  \\
\hline $ac$ & $Q_R$ &  $3( {\bar 3},1,2)$ &  -1  & 0  &  0 \\
\hline $bd$ & $ L $ &  $3(1,2,1)$ &  0   & -1 &  0 \\
\hline $cd$ & $R$  & $3(1,1,2)$ &  0  & 1  & 0  \\
\hline $bc$ &  $H$ & $(1,2,2)$ &  0 & 0 &  0 \\
\hline\hline
$bM$ & $ {F}$ &  $ 6(1,2,1;2_M)$ & 0 & 0 & -1 \\
\hline
$cM$ & $ {G}$ &  $ 6(1,1,2;2_M)$ & 0 & 0 & -1 \\
\hline
\end{tabular}
\end{center}
\caption{\small Chiral spectrum of the MSSM-like left-right symmetric model
discussed in the text. \label{espectrosm2}}
\end{table}

One can easily check that all RR-tadpoles cancel and that, for appropriate choices of closed string moduli, all these branes
preserve  the same \neq1 SUSY as the orientifold background.
The D8-branes $a$ and $d$ give rise to a group 
$SU(3)_c\times  U(1)_{B-L} \times U(1)_{B+L}$,
where the latter $U(1)$ is anomalous and becomes massive as usual through the
$D=4$ Green-Schwarz mechanism.  The D6-branes $b$ and $c$ sit on top
of the horizontal orientifold plane and carry symplectic groups,
in the case at hand one has $USp(2)_L\times USp(2)_R$ corresponding
to a left-right symmetric gauge group $SU(2)_L\times SU(2)_R$. 
Finally, the D8-brane $M$ gives rise to an extra non-Abelian factor $U(2)_M$.

It is easy to compute the chiral spectrum in this model and find
that the chiral spectrum with respect to 
the gauge group $SU(3)_c\times SU(2)_L\times SU(2)_R\times U(1)_{B-L}$ 
has quantum numbers as shown in table \ref{espectrosm2},
i.e. three generations of quarks and leptons. They correspond to open strings 
in between the branes $a$, $d$ and $b$, $c$. In addition there are 
exotic leptons/Higgsses  $6(2_M,2_L)+6(2_M,2_R)$ corresponding to open 
strings in between the brane $M$ and $b$, $c$. However, such exotic matter
can be higgsed away by giving a vev to the symmetric or to the antisymmetric 
fields charged under $U(2)_M$, and which arise from open strings 
stretched between $M$ and its orientifold image $M'$. From the field theory 
point of view, these vevs trigger the gauge symmetry breaking 
$U(2)_M \rightarrow SO(2)_M$ and $U(2)_M \rightarrow SU(2)_M$, 
respectively, after which $6(2_M,2_L)+6(2_M,2_R)$ can acquire a mass.
From a geometric point of view, such process corresponds to a 
D-brane recombination $M + M' \rightarrow \tilde{M}$, and one can see 
that the intersection product of $\tilde{M}$ with the branes $b$ and $c$
vanishes. Finally, such D-brane recombination occurs spontaneously 
in a large region of the complex structure moduli space, as 
can be deduced from the FI-terms of this model.

Notice that $a$, $b$, $c$, $d$ brane structure is analogous to
the MSSM-like model constructed in \cite{cimDESY,cimyuk,ms04}. 
Similar to the latter, 
the MSSM Higgs sector arises from the open strings stretched between the 
D-branes $b$ and $c$, which overlap in the first complex dimension and 
hence give rise to a non-chiral sector of the spectrum.
In addition to the above chiral fields, there are further 
SM singlets which we do not display.

The D-term conditions give in the present example
\beq
 {\tau_2}= {\tau_3} \quad ;\quad  {\tau_1\tau_2\tau_3}
= 9{\tau_1}+6{\tau_2}+6{\tau_3} .
\nonumber
\label{Dtermmodel}
\eeq
Recall that one cannot conclude that the complex structure moduli are
fixed to obey these constraints, since a deviation from these equations
give rise to a FI-terms whose contribution to the vacuum
energy will be cancelled by vevs of chiral matter at the intersections.
In this case gauge symmetry breaking will take place corresponding to brane recombination,
but only some linear combinations of complex structure moduli and matter
scalars are in fact fixed.
On the other hand the presence of the a, X and Y D8-branes give rise to
a superpotential of the form
\beq
W_{D8}\ =\ X_1^a(1-T_2T_3)+ X_2^X(2-T_1T_3)+X_3^Y(2-T_1T_2)
\eeq
where $X_i^{a,X,Y}$ are the geometric $+$ Wilson line open string moduli.
Note that for fixed open string moduli this would fix
the three untwisted K\"ahler moduli to the values
\beq
\re T_1=\ 2 \quad ;\quad  \re T_2=\re T_3 = 1  \quad  ;\quad  \im T_i=0  \  .
\label{fijosm}
\eeq
One can play around with the magnetic fluxes to get other models with larger 
K\"ahler moduli but recall  that, unlike the case of 
moduli fixing through closed string fluxes, the present models correspond to
CFT's and hence we do not need to work in the large volume limit approximation.

\section{Coisotropic branes and Mirror symmetry\label{mirror}}

As emphasized at the introduction, coisotropic branes were first introduced in order to correctly formulate Kontsevich's Homological Mirror Symmetry conjecture. Although we have not made use of this fact anywhere along our discussion, a natural question is how these new type IIA vacua look upon applying mirror symmetry. In particular, if coisotropic D8-branes in {\bf CY}$_3$'s have not been considered before and they are mapped to type IIB D$p$-branes by the mirror map, one may wonder if we have been missing any kind of D-brane in type IIB constructions. This question is particularly meaningful in the case of simple $\T^6/\IZ_2 \times \IZ_2$ orientifolds, where the mirror map is well understood and chiral D-brane vacua have been constructed in both type IIA and type IIB sides.

In the following, we will answer this question for the type IIA $\T^6/\IZ_2 \times \IZ_2$ orientifold considered in this paper, originally introduced in terms of its type I mirror \cite{bl96}. We will see that the type I duals of a coisotropic D8-branes can either be a tilted D5-brane or a magnetized D9-brane, the magnetic flux involved in the latter being rather exotic. Finally, we will make use of the web of T-dualities to show that coisotropic D8-branes are also related to a new class of special Lagrangian D6-branes in $\T^6/\IZ_2 \times \IZ_2$, which have so far not been considered in the construction of $\CN=1$ chiral type IIA vacua.

\subsection{Mirror symmetry to type I}

We will first illustrate how mirror symmetry works by looking at a rather simple example, so let us consider type IIA compactified on $\T^6$ and the BPS, coisotropic D8-brane given by
\beq
\begin{array}{rcl}
\Pi_5 & = & (1,0)_1 \times (\T^2)_2 \times (\T^2)_3\\
F/2\pi & = & dx^2 \wedge dx^3 - dy^2 \wedge dy^3
\end{array}
\label{ccrel}
\eeq
also considered at the beginning of section \ref{coisorientifolds}. If we orientifold this theory by $\Om \CR (-1)^{F_L}$, with $\CR$ given by (\ref{R}), we will introduce an O6-plane on the directions $\{ x^1, x^2, x^3\}$. Hence, in order to map this setup to type I theory compactified on $\T^6$, we need to perform three T-dualities on the directions $\{y^1, y^2, y^3\}$.

\begin{figure}
\epsfysize=9.5cm
\begin{center}
\leavevmode 
\epsffile{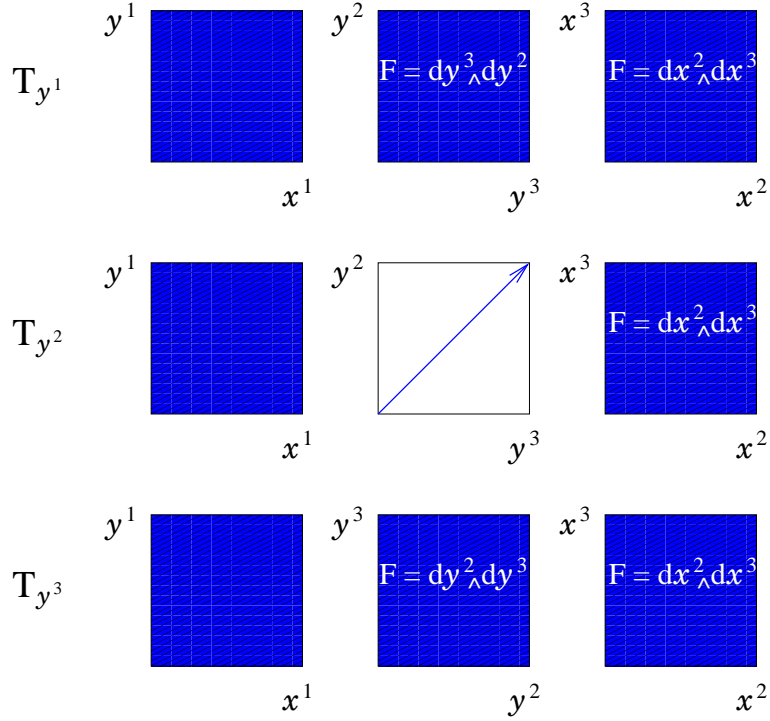}
\end{center}
\caption{Mirror symmetry for the coisotropic D8-brane (\ref{ccrel}).}
\label{mirrorfig}
\end{figure}

How the D8-brane transforms under these three T-dualities is illustrated in figure \ref{mirrorfig}. A T-duality along $y^1$ is trivial and only maps the D8-brane to a D9-brane with the same worldvolume flux $F$. In order to perform the second T-duality we conveniently relabel the coordinates as in the figure, in order not to have components of $F$ which mix different $\T^2$ factors. The T-duality along $y^2$ is then simply deduced from the usual map between torons and branes at angles \cite{bdl96}, obtaining a D8-brane tilted in the $\{y^2, y^3\}$ plane. Finally, applying the same kind of map we recover a D9-brane with worldvolume flux
\beq
F/2\pi \, =\, dx^2 \wedge dx^3 + dy^2 \wedge dy^3.
\label{ccmirror}
\eeq

One of the advantages of this type I picture is that it is easy to work out the orientifold image of such D-brane. Type I theory can be simply seen as type IIB modded out by the world-sheet parity $\Om$, which acts in the D9-brane worldvolume flux as $\Om\, :\, F \rightarrow -F$. Taking the image of (\ref{ccmirror}) and undoing the previous three T-dualities we arrive to the coisotropic D8-brane
\beq
\begin{array}{rcl}
\Pi_5 & = & (-1,0)_1 \times (\T^2)_2 \times (\T^2)_3\\
F/2\pi & = & - dx^2 \wedge dx^3 + dy^2 \wedge dy^3
\end{array}
\label{ccprime}
\eeq
in agreement with (\ref{finalprime}). In addition, one can understand the BPS conditions (\ref{Ftermcc2}) from this mirror picture. Indeed, in the case of type IIB compactified in a Calabi-Yau and in the presence of O9-planes and/or O5-planes, the supersymmetry conditions for a D9-brane read \cite{mmms99}
\beq
\begin{array}{ccrcl}
{\rm D-flatness} & \quad \quad & \frac{1}{3!} \CF^3 & = & \oh J^2 \wedge \CF\\
{\rm F-flatness} & \quad \quad & \CF^{(0,2)} & = & 0
\end{array}
\label{SUSYD9}
\eeq
While the D-flatness condition is trivial for (\ref{ccmirror}), the F-flatness condition requires $F$ to be a (1,1)-form on the D9-brane worldvolume. It is easy to see that, if we parameterize our complex structure moduli as $dz^j\, =\, dx^j + i \mathfrak{t}_j dy^j$ (where $\mathfrak{t}_j$ is now a complex number), this implies that $\mathfrak{t}_2 \cdot \mathfrak{t}_3\, =\, 1$, which is the mirror condition to (\ref{Ftermcc2}).

Type I compactifications on magnetized tori have been considered long ago \cite{typeI}. More recently, magnetic fluxes of the form (\ref{ccmirror}) have appeared in the context of type I theory compactified either on $\T^6$ \cite{am04} or {\bf K3} \cite{kmr06}. Now, if we want to describe the mirror of our type IIA vacua on $\T^6/\IZ_2 \times \IZ_2$ we need to consider type I theory compactified on the $\T^6/\IZ_2 \times \IZ_2$ orbifold with opposite choice of discrete torsion, and embed our magnetized D9-brane in such background. While it may seem that this still is a close relative of the constructions in \cite{am04,kmr06}, there is actually an important difference. Namely, the 2-form flux (\ref{ccmirror}) is cohomologically non-trivial in the case of $\T^6$ and of $\T^2 \times {\bf K3}$, but it becomes trivial for $\T^6/\IZ_2 \times \IZ_2$. In particular, the orbifold group generator (\ref{z2gen1}) acts as $\th\, : \, F \rightarrow -F$ when applied to (\ref{ccmirror}), so $F$ is not even well-defined as a 2-form in $\T^6/\IZ_2 \times \IZ_2$. Of course, this is not a problem for our orbifold construction, because $F$ needs to be well-defined in terms of the D9-branes, and not in terms of the closed string background.\footnote{More precisely, for each D9-brane with worldvolume flux $F$ we will need to add a $\th$-image D9-brane with worldvolume flux $-F$, so that we can quotient our theory by $\IZ_2 \times \IZ_2$.} On the other hand, $F^2$ can be seen as a well-defined, non-trivial representative of a $\T^6/\IZ_2 \times \IZ_2$ cohomology class.

One can get a better intuition of what these facts mean by looking at the induced D-brane charges of our magnetized D9-brane and, in particular, by going back to the original type IIA mirror construction. For instance, the fact that $F$ is trivial in cohomology will imply that the induced charge of D7-brane will vanish, and this statement is  mirror to the fact that our coisotropic D8-brane is wrapping a 5-cycle $\Pi_5$ trivial in homology. More precisely,
\beq
\begin{array}{ccc}
{\rm type\ IIB} & & {\rm type\ IIA}\\
\,[F]\quad {\rm trivial}
& \longleftrightarrow &
\left\{
\begin{array}{c}
\,[\Pi_5]\quad {\rm trivial}\\
\,[\Pi_1^{F^2}]\quad {\rm trivial}
\end{array}
\right.
\\
\left.
\begin{array}{c}
\,[\Pi_6] \quad {\rm non-trivial}\\
\,[F^2]\quad {\rm non-trivial}
\end{array}
\right\}
& \longleftrightarrow &
\,[\Pi_3^F]\quad {\rm Non-trivial} 
\end{array}
\label{Qmap}
\eeq
where $\Pi_6$ is the 6-cycle wrapped by the D9-brane plus its orientifold images. To sum up, there is a conceptual difference between embedding the worldvolume flux (\ref{ccmirror}) in $\T^6/\IZ_2 \times \IZ_2$ with respect to $\T^6$ or $\T^2 \times {\bf K3}$. Such difference is the same than constructing a coisotropic D8-brane in a homologically trivial 5-cycle of a proper ${\bf CY_3}$ rather than in $\T^6$ or in $\T^2 \times {\bf K3}$, where coisotropic branes were already known to exist. 

For our purposes, however, the really interesting property of $\T^6/\IZ_2 \times \IZ_2$ with respect to $\T^6$ or ${\bf K3}$ is that we can 
construct \deq4, \neq1 type I string vacua. Those vacua will contain \deq4 chiral fermions if we introduce non-trivial worldvolume fluxes 
in our D9-branes, even in the case of the more exotic bundles of the form (\ref{ccmirror}). That this is the case was proved in 
Section \ref{chirality} by means of the mirror type IIA picture, but it can also be done directly in type I by generalizing 
the techniques of \cite{bt05} to toroidal orbifolds. We will not attempt to give such general description here, but rather illustrate 
how to construct new \deq4, \neq1  chiral models in type I by describing the mirror of one of the models of the previous section. 

In particular, let us consider the MSSM-like model constructed in table \ref{mssmlr}. If we first consider the Pati-Salam sector of the theory, i.e., branes $a$, $b$ and $c$ we see that only the D-brane stack $a$ changes with respect to the local model in \cite{cimyuk}. In the type I picture, such D8-brane becomes a stack of 10 D9-branes with worldvolume flux
\beq
F_a/2\pi\, =\, \left(\frac{1}{10} (dx^2 \wedge dx^3 + dy^2 \wedge dy^3) +  \frac{3}{10}  (dx^2 \wedge dy^2 - dx^3 \wedge dy^3)  \right) \cdot {\bf 1}_{10}
\label{baryonictypei}
\eeq
This flux will break the initial $U(10)$ gauge group down to $U(1)$, via non-Abelian Wilson lines (see, e.g., \cite{cimmag}), so in fact we will need 4 times this D9-brane content to generate the desired $U(3) \times U(1)$ gauge symmetry and, when embedding our model in $\T^6/\IZ_2 \times \IZ_2$, a total of 80 D9-branes will be needed. Incidentally, notice that in the case of (\ref{baryonictypei}) $\th$ does not map $F_a$ to $-F_a$, but only some components get `projected out'. On the other hand, $F_a^2\, =\, - dx^2 \wedge dy^2 \wedge dx^3 \wedge dy^3$, is well-defined and implies that a D5$_1$-brane charge is induced on the D9-brane.\footnote{Here we are using the usual model building notation, by which D5$_i$ stands for a D5-brane wrapped on the $i^{th}$ complex dimension, and D7$_i$ stands for a D7-brane transverse to the same complex dimension.} Finally, one can check that both supersymmetry conditions in (\ref{SUSYD9}) will be satisfied once that we pick $\mathfrak{t}_2 \cdot \mathfrak{t}_3\, =\, 1$ and ${\rm Area} (\T^2)_2\, =\, {\rm Area} (\T^2)_3$.

The $SU(2)_L \times SU(2)_R$ symmetry of the model arises from the D-branes $b$ and $c$ which, in the type I picture, correspond to a D5$_2$ and a D5$_3$, respectively. Triplication of families can then be understood from the fact that
\beq
\int_{D5_2} \tr F_a \, =\, - \int_{D5_3} \tr F_a\, =\, 3
\label{tripi}
\eeq
and the Higgs sector of the theory will arise from open strings stretching between D5$_b$ and D5$_c$, just like in \cite{cimmag}.

The rest of the model goes as follows. The D-brane $M$ corresponds to $4 \times 18$ anti-D9-branes, which are nevertheless BPS because of the worldvolume flux
\beq
F_M/2\pi\, =\, \left( \frac{1}{2} dx^1 \wedge dy^1 +  \frac{1}{3}  dx^2 \wedge dy^2 + \frac{1}{3} dx^3 \wedge dy^3)  \right) \cdot {\bf 1}_{18}
\label{Mtypei}
\eeq
and a certain choice of K\"ahler moduli. The D-branes $X$ and $Y$ both correspond to $2 \times 2$ D9-branes of the form
\beq
F_{X, Y}/2\pi\, =\,  \left(dx^i \wedge dx^j + \oh dy^i \wedge dy^j\right) \cdot {\bf 1}_2
\label{XYtypei}
\eeq
Finally, we need to add the images of all these D-branes under $\Om$.

Let us also stress that coisotropic D8-branes are not always taken to magnetized D9-branes under the mirror map. Let us consider the D8-brane
\beq
\begin{array}{rcl}
\Pi_5 & = & (0,1)_1 \times (\T^2)_2 \times (\T^2)_3\\
F/2\pi & = & - dx^2 \wedge dy^3 + dx^3 \wedge dy^2
\end{array}
\label{ccc}
\eeq
and again apply three T-dualities on the directions $\{y^1, y^2, y^3\}$. Following the same kind of argument than in figure \ref{mirrorfig}, we obtain a D5-brane wrapping the 2-cycle
\beq
(1,-1)_{(x^2, y^3)} \times (1,1)_{(x^3, y^2)}
\label{cccmirror}
\eeq
and with vanishing worldvolume flux $F$. For the particular choice of complex structure moduli $\mathfrak{t}_2 = \mathfrak{t}_3 = 1$ this becomes a holomorphic two-cycle, given by $z^2\, =\, i z^3$, and which is the required condition for F-flatness \cite{mmms99}. Unlike in the case of a D5$_i$-brane, the holomorphicity of the tilted D5-brane (\ref{cccmirror}) is non-trivial, and this eventually generates a superpotential for the complex structure moduli. In general, one can see that (\ref{cccmirror}) is holomorphic whenever $\mathfrak{t}_2 \cdot \mathfrak{t}_3 = 1$, as we would expect from the mirror BPS conditions. 

\subsection{Back to type IIA}

An interesting property of the $\T^6/\IZ_2 \times \IZ_2$ orientifold considered in this paper is that by performing four T-dualities of the form $\{x^i, y^i, x^j, y^j\}$ we recover the same closed string background. This means that if we construct a BPS D-brane in $\T^6/\IZ_2 \times \IZ_2$ and we apply these four T-dualities, we will obtain a new BPS D-brane in the same orientifold, although the values of the closed string moduli may now be different.

We would like to apply this observation to the coisotropic D8-branes constructed in this paper, in order to see if new BPS D-branes 
can be constructed. Let us then again consider the BPS D8-brane (\ref{ccc}) and apply four T-dualities in the coordinates 
$\{x^1, y^1, x^2, y^2\}$. A similar reasoning than the one carried in figure \ref{mirrorfig} takes us to a D6-brane wrapping 
the 3-cycle\footnote{See \cite{bt05b} for a detailed discussion on similar T-duality computations.}
\beq
\Pi_3 \, =\, (1,0)_{(x^1, y^1)} \times (1,1)_{(x^2,y^3)} \times (1,1)_{(x^3,y^2)} 
\label{exD6}
\eeq
which, despite being a BPS D6-brane, is not one of the usual D6-branes at angles considered in the literature. Indeed, by appropriately 
choosing the closed string moduli of $\T^6/\IZ_2 \times \IZ_2$, (\ref{exD6}) will be related to the O6-planes by means of a 
$SU(3)$ rotation. Away from this point (\ref{exD6}) will no longer be a special Lagrangian, but not because it is not calibrated 
by $\re \Om$, but instead because it is no longer a Lagrangian 3-cycle. That is, just like coisotropic D8-branes, the D6-branes of 
the form (\ref{exD6}) develop non-trivial F-terms via the failure of the BPS conditions (\ref{SUSYD6b}). Given the complexified 
K\"ahler form (\ref{kahler}) it is easy to see that the F-flatness condition is satisfied whenever the untwisted K\"ahler moduli satisfy $T_2 = T_3$. Notice that, unlike the case of coisotropic D8-branes, this is compatible with the limit where all K\"ahler moduli are arbitrarily large.

Notice that (\ref{exD6}) by itself is not well-defined in $\T^6/\IZ_2 \times \IZ_2$, but that we need to include its image under $\th$ in order to have an invariant object under the orbifold group. The image 3-cycle is given by
\beq
\th (\Pi_3) \, =\, (1,0)_{(x^1, y^1)} \times (1,-1)_{(x^2,y^3)} \times (1,-1)_{(x^3,y^2)}
\label{exD6th}
\eeq
which will be a special Lagrangian if (\ref{exD6}) is. Of course, when embedded in $\T^6/\IZ_2 \times \IZ_2$, (\ref{exD6}) and (\ref{exD6th}) should be thought of 3-cycles on the covering space $\T^2 \times {\bf K3}$. In any case, one can see that the homology class of the union of both 3-cycles is given by
\beqa \nonumber
[\Pi_3 \cup \th(\Pi_3)]& = & [(1,0)_1] \otimes [(1,0)_2(1,0)_3 + (0,1)_3 (0,1)_2]\\
& = & [(1,0)_1] \otimes [(1,0)_2(1,0)_3 + (0,1)_2 (0,-1)_3]
\label{exD6union}
\eeqa
which is exactly the D6-brane charge carried by the coisotropic D8-brane (\ref{ccrel}). Notice that both D-branes (\ref{ccrel}) and (\ref{exD6}) are mutually BPS only for the particular choice of K\"ahler moduli $T_2 = T_3 = 1$.

To sum up, by applying four T-dualities on a coisotropic D8-brane we have found a new class of special Lagrangian D6-branes, whose `special' condition is always satisfied but its Lagrangian condition is not. This is quite remarkable, given the fact that we usually encounter the opposite behavior for D6-branes intersecting at angles. Just like in the case of coisotropic D8-branes, this exotic D6-brane carries a D6-brane charge which is a sum of two factorizable 3-cycles (i.e., those of the form $(n^1,m^1)_1 \times (n^2,m^2)_2 \times (n^3,m^3)_3$) while nevertheless being an exact CFT boundary state. It is clear that one can construct many more D6-branes of this kind, so it would be interesting to see which new model building possibilities are opened up by the existence of these less conventional D6-branes.

\section{Conclusions and Outlook \label{tachan}}

Type IIA  \deq4 chiral vacua constructed up to now have been mostly based 
on intersecting D6-branes wrapping special Lagrangian 3-cycles. 
In this paper we have argued that in type IIA Calabi-Yau orientifolds there are
other BPS objects, namely D8-branes wrapping coisotropic 5-cycles,  
which have so far been neglected and which seem to have interesting 
model-building features. These branes wrap 5-cycles in the ${\bf CY}_3$ which are 
trivial in homology, but still are stable BPS objects due to the D6-brane charge 
induced by magnetic fluxes on their worldvolume.
At the intersection of a D8 with another D8 or a D6 chiral fermions 
appear, chirality being generated by a combination of intersecting/magnetized
brane mechanisms. Obviously, this new way of creating \deq4 chirality makes 
coisotropic D8-branes an interesting tool for constructing new 
phenomenologically relevant string vacua.

We have analyzed in detail the case where our ${\bf CY}_3$ is given by a  
$\IZ_2 \times \IZ_2$ type IIA orientifold.
 We have worked out  the chiral spectrum and effective action
   (gauge kinetic function, scalar potential, FI-terms, $U(1)$
anomaly cancellation) for sets of coisotropic D8-branes and
additional D6-branes in this background, and we have also studied the
general form of Yukawa couplings among chiral fields.
These couplings are generated by a combination of wave-function 
overlapping (\`a la Kaluza-Klein) and world-sheet instanton 
contributions, which might furnish new possibilities for the 
flavor dependence of Yukawa couplings in realistic 
compactifications. Thus, it would clearly be interesting to make a 
general exploration of patterns of Yukawa couplings
in realistic models including coisotropic D8-branes.
Another interesting feature of the D8-branes in
this orientifold is that they carry a D6-brane charge which
is not of the factorized form 
{\small (3-cycle) $=$ (1-cycle) $\times$ (1-cycle) $\times$ (1-cycle)}
 in the underlying $\T^2\times \T^2\times \T^2$. 
This opens new model-building possibilities not available to standard
factorized D6-branes in this orientifold. 

The presence of coisotropic D8-branes leads to superpotential couplings 
linear in  open string moduli $X_i$ and bilinear in  untwisted K\"ahler moduli
of the general form $X_i(T_jT_k-n)$.
These couplings could be useful in order to fix K\"ahler moduli in
specific models. In fact in the above orientifold model one could think that
adding an appropriate set of  coisotropic D8-branes one can actually fix
all untwisted K\"ahler moduli. One must be careful, though, since 
in general there may be additional couplings of $X_i$ to other open string
moduli and the minimum of the potential could only fix a linear combination
of K\"ahler and open string moduli, rather than just fixing the K\"ahler
moduli. This is reminiscent of what happens with D-terms from
D6-branes (and also D8-branes) which do not fix the complex structure
moduli but rather a linear combination of those and charged 
chiral scalars. On the other hand it is clear that 
in the presence of other sources of moduli superpotentials 
(like e.g. closed string RR and/or NS fluxes) the existence
of these D8-brane-induced superpotentials could be quite useful
in the general moduli stabilization program. 
In this connection it would be interesting to 
study the effect of RR fluxes, which are known to lead to 
additional K\"ahler-moduli dependent superpotentials,
in models with coisotropic D8-branes. In a more speculative mode,
other possible use of these D8-branes could be to help in 
supersymmetry breaking and the   
up-lifting of the AdS vacua. Indeed, the K\"ahler moduli-dependent
superpotential induced by D8-branes are reminiscent of O'Rafertaigh
superpotentials. An appropriate combination of D8-branes could thus
help in the generation of a (possibly metastable) 
SUSY-breaking vacuum. This possibility should be worth studying.

We have also found that 
the coisotropic D8-branes that we construct in the
$\IZ_2 \times \IZ_2$ orientifold have well defined type I
T-duals. They correspond either to certain tilted D5-branes or
D9-branes with off-diagonal fluxes. One could also T-dualize
to the case of type IIB with O3/O7-planes. This could be interesting
in order to make contact with the kind of models
with closed string RR and NS fluxes which have been considered in
order to fix all moduli in IIB orientifolds. 
On the other hand we have found that four T-dualities
convert certain  coisotropic D8-branes into a new class of BPS D6-branes
not previously considered in the model-building literature.
 Interestingly enough,
unlike the standard factorized  D6-branes considered up to now,
these D6-branes give rise to non-trivial F-term
conditions of the type $T_i = T_j$ in these orientifolds and
have also non-factorized RR D6-charge. Consideration of 
this new class of D6-branes could also be interesting 
from the model-building point of view and is at present under 
study.

As an illustration of the model-building properties of 
coisotropic D8-branes we have presented two examples with
a chiral spectrum very close to that of the MSSM,
which are phenomenologically interesting in their own right. In the
first model the MSSM branes are made of D6-branes and a stack of 
D8-branes helps in cancelling all RR tadpoles in a more efficient way 
that D6-branes would do. The second 
example provided is a D8-brane relative of the left-right 
symmetric D6-brane model constructed in \cite{cimDESY,cimyuk,ms04}.
A general statistical analysis of how easy it is to find consistent
models of this second type and how many have an MSSM-like spectrum  in the present
$\IZ_2 \times \IZ_2$ orientifold constructions  was performed in
\cite{gbhlw}, and more recently in \cite{dt}. It is clear that 
if we take into account the freedom of adding coisotropic D8-branes 
in these  constructions new possibilities (like the two 
MSSM-like examples discussed in this paper) will appear.
 Thus, presumably the landscape of MSSM-like models within the 
$\T^6/\IZ_2 \times \IZ_2$ orientifold will be wider than we previously thought.

\bigskip

\bigskip

\centerline{\bf \large Acknowledgments}

\bigskip

We would like to thank R.~Blumenhagen, P.G.~C\'amara, R.~Empar\'an, S.~Theisen, and specially A.~Uranga for useful discussions. L.~E.~I. and F.~M. would also like to thank CERN PH-TH division for hospitality during the completion of the paper. This work has been partially supported by the European Commission under the RTN European Program MRTN-CT-2004-503369, the Comunidad Aut\'onoma de Madrid (proyecto HEPHACOS; P-ESP-00346) and the CICYT. The work of F.M. is supported by the European Network ``Constituents, Fundamental Forces and Symmetries of the Universe", under the contract MRTN-CT-2004-005104.

\newpage

\appendix

\section{Formal definition of coisotropic branes\label{app}}

For completeness, in this appendix we state the formal definition of coisotropic A-brane, as originally described by Kapustin and Orlov. Our intention is not to give a full report on this subject, but rather to provide the basic definitions that connect our results with the topological A-brane literature. For a more detailed discussion we refer the reader to the original reference \cite{ko01} and related literature \cite{op03,ko03,Oh03,kl03}.

Let us consider a symplectic manifold $\CM$, so that it contains a globally well-defined, non-degenerate, closed two-form $J$. Given a vector field $X$ on $\CM$ we can map it to a 1-form $\xi_X$ by the usual contraction of indices
\beq
\xi_X\, =\, \iota_X J
\label{Jmap}
\eeq
and, because of the properties of $J$, this map is an isomorphism. This means that we can define an inverse, $J^{-1}$, which maps 1-forms to vectors and, in general, covariant indices to contravariant indices.

Let us now consider a submanifold $\CN \subset \CM$ and define
\beq
{\rm Ann}\, T_\CN \, = \, \{\xi \in T_\CM^{\, *}\ |\ \xi_\rho X^\rho = 0,\, \forall X \in T_\CN \}
\label{Ann}
\eeq
where $T_\CN$ is the tangent bundle of $\CN$. We then have the following definitions:
\begin{itemize}

\item{If $J$ takes $T_\CN$ inside Ann $T_\CN$\ $\Rightarrow$\ $\CN$ is an isotropic submanifold}

\item{If $J^{-1}$ takes Ann $T_\CN$ inside $T_\CN$\ $\Rightarrow$\ $\CN$ is a coisotropic submanifold}

\item{If both conditions are satisfied\ $\Rightarrow$\ $\CN$ is a Lagrangian submanifold}

\end{itemize}

One can easily apply these conditions to the examples of coisotropic submanifolds given in the main text. If, for instance, we consider the D8-brane (\ref{cc}) then one has
\beq
\begin{array}{rcl}
\Pi_5 & = & (1,0)_1 \times (\T^2)_2 \times (\T^2)_3\\
T_{\Pi_5} & = & \langle X^1, X^2, Y^2, X^3, Y^3\rangle \\
{\rm Ann}\, T_{\Pi_5} & = & \langle dy^1\rangle
\end{array}
\label{excoiso}
\eeq
and so, because $J$ is given by (\ref{kahler}), $J({\rm Ann}\, T_{\Pi_5}) = \langle X^1\rangle  \subset  T_{\Pi_5}$ and so the manifold is coisotropic. Notice that embedding $\Pi_5$ in $\T^6/\IZ_2 \times \IZ_2$ does not change this conclusion.

While the definitions above are rather simple, in order to describe a coisotropic A-brane it is more convenient to use an alternative, equivalent definition of coisotropic submanifold. Let us first define the symplectic orthogonal bundle of $\CN$ as
\beq
(T_\CN)^J \, = \, \{Y \in T_\CM \ | \ Y^\rho J_{\rho\sigma} X^\sigma = 0, \, \forall X \in T_\CN \}
\label{ort}
\eeq
and then define a coisotropic submanifold whenever $(T_\CN)^J \subseteq T_\CN$. Because $J$ is non-degenerate, this inclusion 
can only be an equality when $\CN$ is middle-dimensional with respect to $\CM$, and in that case we are dealing with a Lagrangian 
submanifold. For the same reason, we can define the quotient bundle $T_\CN/(T_\CN)^J$. In our example above, it is easy to see that $(T_\CN)^J = \langle X^1 \rangle$, and hence $T_\CN/(T_\CN)^J = T_{(\T^2)_2 \times (\T^2)_3}$.

While the first supersymmetry condition for an A-brane is that it wraps a coisotropic submanifold, the other two involve the worldvolume flux $\CF$ and they read:
\begin{itemize}

\item{$\CF$ has no components on $(T_\CN)^J$, so it can be defined as a 2-form living in $T_\CN/(T_\CN)^J$}

\item{$J^{-1} \CF$ defines a complex structure on $T_\CN/(T_\CN)^J$}

\end{itemize}
Finally, from these conditions one can show that $T_\CN/(T_\CN)^J$ has even complex dimension, and that if ${\dim}\, (\CM) = 2n$, then  ${\dim}\, (\CN) = n + 2k$, $k \in \IN$.

Let us check that those conditions are also satisfied for our canonical example (\ref{cc}). It is clear that $\CF = B|_{\Pi_5} + 2\pi\a' F$ has no components on $(T_{\Pi_5})^J$, so it only remains to check that $J^{-1} \CF$ is a complex structure in $T_\CN/(T_\CN)^J = T_{(\T^2)_2 \times (\T^2)_3}$. Applying the definitions above we have that
\beq
J^{-1}\CF\, =\,
\left(
\begin{array}{cccc}
-\frac{\im T_2}{\re T_2} & 0 & 0 & -\frac{1}{\re T_2} \\
0 & -\frac{\im T_2}{\re T_2} & -\frac{1}{\re T_2} & 0 \\
0 & -\frac{1}{\re T_3} &  -\frac{\im T_3}{\re T_3} & 0 \\
-\frac{1}{\re T_3} & 0 & 0 & -\frac{\im T_3}{\re T_3}
\end{array}
\right)
\label{complex}
\eeq
and so it is easy to see that $(J^{-1}\CF)^2 = -{\bf 1}_4$ if and only if $\im T_2T_3 = 0$ and $\re T_2T_3 = 1$, which is precisely the F-flatness condition in (\ref{Ftermcc}).

\end{document}